\documentclass[prb,preprint]{revtex4}
\usepackage{graphicx}
\makeatletter
\usepackage{epsfig}
\def\@dotsep{4.5}
\begin{document}

\title{Theory of Direct Scattering, Trapping and Desorption in Atom-Surface Collisions}

\author{Guoqing Fan and J. R. Manson}
\affiliation{ Department of Physics and Astronomy, Clemson University, Clemson, SC, 29634}
\email{jmanson@clemson.edu} %optional
\date{\today}

\begin{abstract}

When gas atoms or molecules collide with clean and ordered surfaces, under many  circumstances the
energy-resolved scattering spectra exhibit two clearly distinct features due  to direct scattering and to
trapping in the physisorption  well with subsequent desorption.  James Clerk Maxwell is credited with being the
first to describe this situation by invoking the simple assumption that when an impinging gas beam is scattered from a surface it can be
divided into a part that exchanges no energy and specularly reflects and another part that equilibrates or
accommodates completely and then desorbs with an equilibrium distribution.   In this paper a scattering theory
is developed, using an iterative algorithm and classical mechanics for the collision process, that describes
both direct scattering and trapping-desorption of the incident beam.  The initially trapped fraction of
particles can be followed as they continue to make further interactions with the surface until they are all
eventually promoted back into the positive energy continuum and leave the surface region.  Consequently, this
theory allows a rigorous test of the Maxwell assumption and determines the conditions under which it is valid.
The theory also gives quantitative explanations of recent experimental measurements which exhibit both a direct scattering
contribution and a trapping-desorption fraction in the energy-resolved spectra.
\end{abstract}

\pacs{34.35.+a,34.50.-s,82.20.Rp}

\maketitle

\section{Introduction}

Trapping and sticking are important processes that occur in gas-surface interactions. Sticking is generally
associated with strong chemical bonding to the surface and once bound the stuck particle does not desorb.
Trapping, on the other hand, is associated with the physisorption potential well created by the relatively weak Van der Walls
potential and in many circumstances trapped gas particles will desorb after a period of residing in the well
near the surface.

Maxwell, in his studies of gas-surface interactions, is credited with being the first to invoke the simple assumption that a
gas impinging on a surface is scattered into two fractions, one that reflects specularly and exchanges no energy and
the other that equilibrates or accommodates completely and desorbs with an equilibrium
distribution.~\cite{Maxwell} This idea was taken up early in the twentieth century by Knudsen who introduced the
concept of the term ``coefficient of thermal accommodation" to measure the efficiency of energy exchange at the
interface between a gas and a surface and developed a theoretical framework in which to describe
it.~\cite{Knudsen} Since this early work it has become standard to assume in gas-surface collisions that the
fraction of the incident gas beam that is trapped and subsequently desorbed leaves the surface in an equilibrium
distribution, i.e., its accommodation coefficient is assumed to be unity and its distribution function is the
Knudsen flux. This assumption of Maxwell appears to be very useful because it appears to explain qualitatively
experimental results measured under  a wide range of different conditions, however, such an assumption has never
been adequately verified theoretically.

In fact, it is clear that the first part of the Maxwell assumption, i.e., that the direct scattering
contribution is elastic can hold strictly true only in the case of quantum mechanical conditions which implies
low energies, small temperatures and small mass ratios.  For classical scattering, which is the regime of large
energies, high temperatures and large mass ratios the direct scattering contribution leaves the surface as a
distribution over a range of energies whose average is typically smaller than the energy of the incident beam,
but may be larger in the case of high surface
temperatures.~\cite{Janda-PRL-79,Hurst-JCP-85,Tully-JCP-90,Sibener-JCP-03,Morris-JCP-2003} However, the second
part of the Maxwell assumption, i.e., that the trapping-desorption fraction leaves the surface in equilibrium,
is often still today used to explain experiments.~\cite{Sibener-JCP-03,Morris-JCP-2003}

The purpose of this paper is twofold.  First, we  test the Maxwell assumption of equilibrium for the
trapping-desorption fraction  using realistic calculations for a simple model of the gas-surface potential and
determine when such an assumption is valid.  Second, we demonstrate that a model of the interaction potential
that retains the basic elements necessary for trapping, when combined with a calculation that contains correct
statistical mechanics, can explain modern high-precision energy-resolved scattering measurements. The gas projectile
is taken to be an atom and its interaction potential is taken to be an attractive square well with a strongly
repulsive surface barrier. The calculations are carried out using classical mechanics, which is justified for
many systems of interest in rarefied surface dynamics. A classical treatment means that the results will
describe heavy mass atoms at higher energies and surfaces at high temperatures where quantum mechanical effects
are not dominant. The use of an attractive square well to approximate the slowly-varying Van der Waals potential
is also reasonable when used with a classical calculation since it gives a good description of the two primary
effects upon entering the well which are an increase in energy and a refraction of the atom to steeper angles
toward the surface. The authors have earlier presented results using a similar model for
one-dimensional scattering. This work extends those earlier results to the much more realistic and more
complicated case of fully three-dimensional scattering.~\cite{Guoqing}

Once a beam of incoming atoms interacts with the surface, a fraction will be directly scattered while the
remainder will be trapped in the potential well. Of the trapped fraction, some will lose sufficient energy to be
actually trapped in the well with negative total energy while others, even though they have positive total
energy, will scatter at angles sufficiently close to grazing that they will be deflected back towards the
surface by the attractive part of the well. This latter positive energy part of the trapped particles is often
called the chattering fraction. The trapped portion of the incident beam particles will continue to have interactions with
the surface and with each subsequent interaction some will receive enough energy and will be projected
sufficiently close to the surface normal that they can escape, while the remainder will continue to be trapped.
Eventually, in a closed system, all particles initially trapped will ultimately desorb from the surface,
although for low temperatures and deep potential wells this could take a very large time.

For the model considered here, through an iterative algorithm, all of the trapped particles are followed as they
continue to have collisions with the surface, and at each iteration the negative energy fraction, the chattering
fraction, and that fraction which is desorbed is recalculated. In this manner the energy distribution of the
slowly diminishing trapped particle fraction as well as the energy and angular distribution of the desorbed (or
scattered) particles can be followed, and a trapping time can be calculated. By following the initial direct
scattering and the sum of all the subsequently desorbed particles the approach to equilibrium of the
trapping-desorption fraction can be monitored.

What is determined  is that for shallow potential wells and large surface temperatures,
conditions for which little trapping is expected, the desorbed fraction leaves the surface very quickly and does
not at all resemble an equilibrium Knudsen flux. However, for deeper wells and lower temperatures when the
majority of the incident particles are trapped, it is found that the energy distribution of the trapped
fraction rather quickly saturates to a stable functional form while the total number of trapped particles slowly
diminishes. As a function of initial conditions favorable to trapping, e.g., low incident beam energy, deep
wells and low surface temperatures, the approach of the scattered particles towards an equilibrium distribution
is followed. We find that in many circumstances, the energy distribution of the scattered particles readily
approaches equilibrium shape even for trapping times that are relatively short. The approach of the angular
distribution to equilibrium shape, which is a Knudsen cosine distribution independent of azimuthal angle, occurs
more slowly and only for very large trapping times. Thus this work provides a real prediction for the conditions
under which the Maxwell assumption of equilibrium for the trapping-desorption fraction can  be applied with
reasonable accuracy.

However, the mere fact that these calculations can indicate the conditions  under which the  trapping-desorption
fraction may appear as a nearly equilibrium distribution is not sufficient to demonstrate that such conditions
are realistic.  In order to be convincing, calculations with the same potential model should be capable of
explaining  real experimental measurements.  To demonstrate this ability, we have chosen to compare our calculations
with recent high-precision energy-resolved data for the scattering of Ar atoms from a well-ordered monolayer of the
polymer 1-decanethiol adsorbed on Au(001).~\cite{Sibener-JCP-03}  This experiment showed that for well-defined
beams of Ar incident over a large range of energies and angles the scattered distributions could be described by
a combination of two features, a direct scattering fraction and a trapping-desorption fraction that was nearly
in equilibrium with the surface.  Our calculations provide an excellent description of both contributions of the
scattered spectra and produce an estimate of the average physisorption well depth of the interaction potential.

In the remainder of this paper the theory is fully developed together with a description of the iteration method
in the following Section~\ref{theory}. In Sec.~\ref{result}, a number of  calculated results describing the
approach of the trapping-desorption fraction towards equilibrium are shown and discussed.  In
Sec.~\ref{sibener}, calculations are compared with the experimental data of Ref.~[\cite{Sibener-JCP-03}].
Conclusions are discussed in Sec.~\ref{conclusion}.

\section{Theory}\label{theory}

A convenient way of describing a surface scattering event  is through a
differential reflection coefficient, written as ${dR({\bf p}_f, {\bf p}_i)} / {dE_f d\Omega _f}$ which gives the
fraction of an incident beam of momentum ${\bf p}_i$ that is scattered into the small energy interval and small
solid angle in the direction of the scattered momentum ${\bf p}_f$. The differential reflection coefficient
obeys the unitarity condition which assures that the number of particles scattered equals the number incident on
the surface
\begin{equation} \label{unitarity}
\int_0^\infty dE_f \int d\Omega _f \frac{dR({\bf p}_f,{\bf p}_i)}{dE_f d\Omega _f} = 1
\end {equation}

The interaction potential is specified by a vibrating repulsive wall with a square physisorption well in front of depth $D$ and  width $b$, where the actual length of $b$ plays
a role in calculating the trapping times but is unimportant for all other calculations as long as it is larger
than the selvage region
containing the vibrational corrugations
 of the repulsive potential. It is assumed that the  exchange of energy and momentum with
the surface occurs only in collisions with the repulsive wall, while all trapped particles collide with a
stationary wall at the front of the well positioned at $z=b$ that simply reflects specularly. This means that
for the purpose of calculating the resulting distribution after a collision the differential reflection
coefficient is calculated using momenta that include the well depth in the normal component. For example, a
particle that would have momentum ${\bf p}_q$ outside the well has the momentum ${\bf p}^\prime_q$ inside where
the two differ in that the energy associated with perpendicular motion is increased by the well depth
\begin{equation}  \label{refr1a}
p'^2_{qz} = p^2_{qz} + 2mD .
\end{equation}
where $m$ is the atom mass. Similarly, an atom incident from asymptotically far away with polar angle $\theta_i$
with respect to the surface normal will be refracted inside the well into the angle
\begin{equation} \label{refra}
\cos \theta '_i = \sqrt { \frac {E_i \cos ^2 \theta _i + D}{E_i + D}}.
\end{equation}
The relationship between the differential reflection coefficients inside and outside of the well for the
fraction which escapes is given by a simple Jacobian that is calculated from Eqs.~(\ref{refr1a})
and~(\ref{refra}).

Within this model, the trajectory of a given atom consists of successive collisions with the surface. The
incoming beam first enters the well and then proceeds to have a first collision with the repulsive wall that scatters  into a
distribution of energies and angles dictated by the differential reflection coefficient. Some of these scattered
particles have sufficient energy and small enough polar angles to escape out of the well, this is the direct
scattering portion. The remaining particles are trapped, they are specularly reflected by the front face of the
potential well, and then they travel back to the repulsive potential where they suffer a second collision. This
process repeats multiple times until all of the initially trapped atoms eventually escape the confines of the
potential well.

Based on a zeroth order differential reflection coefficient ${dR^0({\bf p}_f, {\bf p}_i)} / {dE_f d\Omega _f}$
which, for each collision with the repulsive potential, gives the probability of scattering from momentum state
$ {\bf p}_i$  to $ {\bf p}_f$ the total differential reflection coefficient after $n$ such collisions can be
written schematically as
%%XXXXXXXXXXXXXXXXXXXXXXXXXXXXXXXXXXXXXXXXXXXXXXXXXXXXXXXXXX
%%
%%
\begin{eqnarray} \label{M1}
\frac{dR^n({\bf p}_f,{\bf p}_i)}{d  {E}_f d \Omega_f}  ~=~ \frac{dR^0({\bf p}_f,{\bf p}_i)}{d  {E}_f d \Omega_f}
~+~ \int d E_b d \Omega_b ~ \frac{dR^0({\bf p}_f,{\bf p}_b)}{d  {E}_f d \Omega_f} ~ \frac{dR^0({\bf p}_b,{\bf
p}_i)}{d  {E}_b d \Omega_b}
\\ \nonumber
~+~ \int d E_b d \Omega_b ~ \frac{dR^0({\bf p}_f,{\bf p}_b)}{d  {E}_f d \Omega_f} ~ \frac{dR^1({\bf p}_b,{\bf
p}_i)}{d  {E}_b d \Omega_b} ~+~ \ldots
\\ \nonumber
~+~ \int d E_b d \Omega_b ~ \frac{dR^0({\bf p}_f,{\bf p}_b)}{d  {E}_f d \Omega_f} ~ \frac{dR^{n-1}({\bf
p}_b,{\bf p}_i)}{d  {E}_b d \Omega_b} ~~~,
\end {eqnarray}
%%
%%
%%XXXXXXXXXXXXXXXXXXXXXXXXXXXXXXXXXXXXXXXXXXXXXXXXXXXXXXXXXX
where the intermediate integrations in the higher order terms are carried out only over those energies and
angles that pertain to particles trapped in the bound states.

Such a procedure lends itself  to an iterative formulation in which the scattered distribution remaining
in the well after the last collision becomes the source for the next collision. The angular and energy space
within the well is divided into bins sufficiently small so as to obtain good numerical precision and it is
necessary to keep track separately of the three different types of trajectories, i.e., the trapped particles
with negative total energy (the trapping fraction), the trapped particles with positive energy (the chattering
fraction) and those that escape at each iteration (the trapping-desorption fraction).

An explicit mathematical description of how this is accomplished is as follows. After the $n$th ($n\geq1$)
iteration, the differential reflection coefficient inside the potential well ${dR^n({\bf p}'_f,{\bf
p}'_i)}/{dE'_f d\Omega '_f}$ is:
\begin{eqnarray}\label{drc}
\frac{dR^n({\bf p}'_f,{\bf p}'_i)}{dE'_f d\Omega '_f}& = & \left\{
\begin{array}{cc} \displaystyle \frac{dR^{n-1}({\bf p}'_f,{\bf
p}'_i)}{dE'_f d\Omega '_f} + \frac{dR_{Con}^n({\bf p}'_f,{\bf p}'_i)}{dE'_f d\Omega '_f} &~;~ {E'_f
> D},{ 0< \theta '_f < \theta '_{fc} } \\
\displaystyle \frac{dP^n({\bf p}'_f, {\bf p}'_i)}{dE'_f d\Omega'_f} &~;~ {otherwise}
\end{array} \right.
~,
\end{eqnarray}
where the upper line of the left hand side of Eq.~(\ref{drc}), labeled with the conditions ${E'_f > D}$ and ${ 0< \theta '_f < \theta
'_{fc} } $,  gives the intensity scattered into the continuum states after  $n$ iterations and consists of the
contribution that was already in the continuum state after $n-1$ iterations plus the fraction contributed to the
continuum by the $n$th iteration.
The critical angle for reflection of particles in the positive energy chattering fraction from the front of the well is $\theta'_{fc} $ which is dependent on energy and given by an equation similar to Eq.~(\ref{refra}).
The fraction that remains trapped in the well is  divided into the sum of the
positive energy chattering fraction and the
 negative energy trapped fraction, denoted respectively by the subscripts $C$ and $T$, according to
\begin{equation}
\frac{dP^n({\bf p}'_f, {\bf p}'_i)}{dE'_f d\Omega'_f}= \frac{dR_C^n({\bf p}'_f,{\bf p}'_i)}{dE'_f d\Omega
'_f}+\frac{dR_T^n({\bf p}'_f,{\bf p}'_i)}{dE'_f d\Omega '_f}
\end{equation}
The positive energy chattering and negative energy trapped fractions are further divided as follows:
\begin{equation}
\frac{dR_C^n({\bf p}'_f, {\bf p}'_i)}{dE'_f d\Omega '_f} = \left\{ {[1- N({\bf p}'_f; \theta'_f)]
\frac{dR_C^{n-1}({\bf p}'_f,{\bf p}'_i)}{dE'_f d\Omega '_f} + \frac{dR_{IC(C)}^n({\bf p}'_f,{\bf p}'_i)}{dE'_f
d\Omega '_f}+ \frac{dR_{IT(C)}^n({\bf p}'_f,{\bf p}'_i)}{dE'_f d\Omega '_f}}  \right\}
\frac{1}{\mathcal{N}^n}
~,
\end{equation}
and
\begin{equation}
\frac{dR_T^n({\bf p}'_f,{\bf p}'_i)}{dE'_f d\Omega '_f} = \left\{  { [1- N({\bf p}'_f; \theta'_f)]
\frac{dR_T^{n-1}({\bf p}'_f,{\bf p}'_i)}{dE'_f d\Omega '_f} + \frac{dR_{IC(T)}^n({\bf p}'_f,{\bf p}'_i)}{dE'_f
d\Omega '_f}+ \frac{dR_{IT(T)}^n({\bf p}'_f,{\bf p}'_i)}{dE'_f d\Omega '_f}}  \right\}
\frac{1}{\mathcal{N}^n}
~.
\end{equation}
In the above equations the factor $ [1-N({\bf p}'_f; \theta'_f)]$ where $ N({\bf p}'_f; \theta'_f) $  is the
ratio of normal velocity to that of the maximum normal velocity of all bound states, given by
\begin{equation}  \label{times}
N({\bf p}; \theta )= \frac {p \cos \theta}{P^{Max}_z} ~,
\end{equation}
where $P^{Max}_z$ is the largest normal momentum component of all the trapped particles and  $p \cos \theta$ is
the momentum component in the $z$ direction for any other trapped particle. The term multiplied by this factor takes
account of the fact that the slower particles collide less often than the faster particles. The
$\mathcal{N}^n$ is a normalization coefficient chosen such that ${dR^n({\bf p}'_f,{\bf p}'_i)}/{dE'_f d\Omega
'_f}$ is normalized  as in Eq.~(\ref{unitarity}).
The intermediate differential reflection coefficients for the chattering and negative energy trapped fractions
at each iteration are given by
\begin{equation}\label{drcchat}
\frac{dR_{IC(X)}^n({\bf p}'_f,{\bf p}'_i)}{dE'_f d\Omega '_f}= \int _D^\infty {dE''_q \int_{\theta
'_{fc}}^{\frac{\pi}{2}} {d\theta ''_q } \int_0^{2\pi } {d\phi ''_q } } \frac{dR^0({\bf p}'_f,{\bf p}''_q)}{dE'_f
d\Omega '_f} N({\bf p}''_q; \theta ''_q )\frac{dR_C^{n-1}({\bf p}''_q, {\bf p}'_i)}{dE''_q d\Omega''_q} ~,
\end{equation}
and
\begin{equation}\label{drctrap}
\frac{dR_{IT(X)}^n({\bf p}'_f,{\bf p}'_i)}{dE'_f d\Omega '_f}= \int_0^D {dE''_q} \int_0^{\frac{\pi }{2}}
{d\theta ''_q } \int_0^{2\pi } {d\phi ''_q } \frac{dR^0({\bf p}'_f,{\bf p}''_q)}{dE'_f d\Omega '_f} N({\bf
p}''_q; \theta ''_q ) \frac{dR_T^{n-1}({\bf p}''_q, {\bf p}'_i)}{dE''_q d\Omega''_q} ~,
\end{equation}
where the symbol $X$ can stand for any one of the three possibilities: $C$ for the chattering fraction, $T$ for the negative energy trapped
fraction or $Con$ for the fraction that goes into the continuum.  For example, $ {dR_{IT(C)}^n({\bf p}'_f,{\bf
p}'_i)} / {dE'_f d\Omega '_f}$ is the intermediate differential reflection coefficient giving the probability
during the $n$th iteration that a particle will make a transition from the negative energy trapped fraction to
the chattering fraction. Finally, the contribution to the continuum states  in the total differential
reflection coefficient of Eq.~(\ref{drc}) coming from the $n$th iteration is
\begin{equation}
\frac{dR_{Con}^n({\bf p}'_f,{\bf p}'_i)}{dE'_f d\Omega '_f} =    \frac{1}{\mathcal{N}^n}  ~ \frac{dR_{IC(Con)}^n({\bf p}'_f,{\bf
p}'_i)}{dE'_f d\Omega '_f   }   ~+~
 \frac{1}{\mathcal{N}^n} ~ \frac{dR_{IT(Con)}^n({\bf p}'_f,{\bf p}'_i)}{dE'_f d\Omega '_f  }
 ~.
\end{equation}

At the end of  $n$ iterations the fraction of all incident particles that remain trapped in the positive energy
chattering states is
\begin{equation}  \label{A1}
P_{C}^n = \int _D^\infty {dE'_f} \int_{\theta '_{fc} }^{\frac{\pi}{2}} {d\theta '_f} \int _0^{2\pi } {d\phi '_f}
\frac{dP_T^n({\bf p}'_f, {\bf p}'_i)}{dE'_f d\Omega'_f} ~,
\end{equation}
while the fraction trapped with negative total energies is
\begin{equation}  \label{A2}
P_T^n =\int_0^D {dE'_f} \int_0^{\frac{\pi }{2}} {d\theta '_f } \int_0^{2\pi } {d\phi '_f } \frac{dP_T^n({\bf
p}'_f, {\bf p}'_i)}{dE'_f d\Omega'_f} ~.
\end{equation}
Thus the total trapped fraction after $n$ iterations is
\begin{equation}
P^n=P_{C}^n + P_T^n ~.
\end{equation}
The fraction escaping into the continuum state after $n$ iterations is
\begin{equation}  \label{A3}
P_{Con}^n = \int _D^\infty {dE'_f} \int _0^{\theta '_{fc} } {d\theta '_f }\int _0^{2\pi} {d\phi '_f } \frac{dR^n({\bf
p}'_f,{\bf p}'_i)}{dE'_f d\Omega '_f} ~,
\end{equation}
and the unitarity condition assures that the total number of particles is conserved
\begin{equation}
P_{C}^n + P_T^n + P_{Con}^n= 1 ~.
\end{equation}

The major numerical operation in this procedure are the two volume integrals associated with evaluating the intermediate differential reflection coefficients of Eqs.~(\ref{drcchat}) and ~(\ref{drctrap}) and the unitarity summations of Eqs.~(\ref{A1}),~(\ref{A2}) and ~(\ref{A3}), which taken together amounts to a 6-dimensional integral.  The angular integrations are carried out using Gauss-Legendre quadratures and the energy integrals use Gauss-Laguerre quadratures.  Because  classical differential reflection coefficients are positive definite and typically tend to consist of a single broad peak or a small number of such peaks in both the energy and angular variables, Gauss quadratures are ideally suited for these integrals.

The only remaining element of this procedure that needs to be specified is the zeroth order  differential
reflection coefficient. There are a number of choices that have been used in the past to describe classical
mechanical collisions of atoms with vibrating surfaces.~\cite{dis1,Brako-Newns,Manson-Muis} The simplest of
these, and the most appropriate for the present calculations, is the differential reflection coefficient for an
atomic projectile colliding with a surface of discrete scattering centers of mass $M$ whose initial momenta are
distributed in an equilibrium distribution at temperature $T_S$. This is given by~\cite{dis1,dis2,dis3}:
\begin{equation} \label{dis}
\frac{dR^0({\bf p}_f,{\bf p}_i)}{dE_f d\Omega _f} = \frac {m^2 \left| {{\bf p}_f} \right|} {8\pi ^3 \hbar ^4
p_{iz} } \left| {\tau _{fi} } \right|^2 \left(\frac{\pi }{k_B T_S \Delta E_0 }\right)^{1/2} \exp \left\{ -
\frac{(E_f - E_i + \Delta E_0 )^2 }{4k_B T_S \Delta E_0 } \right\} ~,
\end {equation}
where $\Delta E_0 = ({\bf p}_f - {\bf p}_i )^2 /2M $ is the recoil energy, $p_{iz}$ is the $z$ component of the
incident momentum, $k_B$ is the Boltzman constant, $\left| {\tau _{fi} } \right|^2$ is the form factor of the
scattering center which depends on the interaction potential. To lowest order, the amplitude $ {\tau _{fi}
} $ is identified as the transition matrix element of the elastic interaction potential extended off the
energy shell~\cite{Celli-Himes-Manson}, however, for this work we use the value appropriate for hard sphere
scattering which is a constant. The differential reflection coefficient of Eq.~(\ref{dis}) can be obtained from
a purely classical calculation or from a quantum mechanical formulation in which the classical limit is
extracted. In the case of a completely classical derivation the constant $\hbar$ is unspecified except for its
dimensions of action, whereas quantum derivations identify $\hbar$ as Planck's constant divided by $2 \pi$.

To obtain the trapping time $\tau$ a variety of methods can be used, but we have found that the most convenient
is to first calculate the average speed normal to the surface for the trapped particles after each iteration.
The time for that iteration is then determined as that required to travel the distance $2b$ from the repulsive
wall to the front of the well and back.

The average normal speed for the positive energy trapped fraction is
\begin{equation} \label{spav1}
<v_n>_{C} = \frac{1}{P_{C}^n} {\int _D^\infty {dE'_f} \int_{\theta '_{fc} }^{\frac{\pi}{2}} {d\theta '_f} \int
_0^{2\pi } {d\phi '_f} \sqrt{\frac{2E'_f}{m}} \cos(\theta '_f) \frac{dP_T^n({\bf p}'_f, {\bf p}'_i)}{dE'_f
d\Omega'_f}},
\end{equation}
and for the negative energy trapped fraction it is
\begin{equation} \label{spav2}
<v_n>_T = \frac{1}{P_T^n} {\int_0^D {dE'_f} \int_0^{\frac{\pi }{2}} {d\theta '_f } \int_0^{2\pi } {d\phi '_f}
\sqrt{\frac{2E'_f}{m}} \cos(\theta '_f) \frac{dP_T^n({\bf p}'_f, {\bf p}'_i)}{dE'_f d\Omega'_f}},
\end{equation}
The total trapping time is then given by summing the times at each iteration
\begin{equation} \label{time1}
\tau= 2 b \sum_n  \left(  \frac{1}{<v>^n_T} \frac {P^n_T}{P^0} + \frac{1}{<v>^n_{C}}\frac {P^n_{C}}{P^0}
\right) ~=~ \tau_T ~+~ \tau_C ~,
\end {equation}
where $P^0$ is the fraction of initially trapped particles. The actual definition of trapping time used in this
work is the time required for the fraction of trapped particles remaining in the well to be reduced to one
 percent of the number of incident atoms.

The method of calculation of the average trapping time is clearly not unique and we have evaluated it several
ways using the trapped fraction probabilities as in Eqs.~(\ref{spav1}) and ~(\ref{spav2}) above. For example,  instead of determining the average speed one
can use the root mean square normal speed, or find the average time directly by obtaining the average of
$2b/v_z $ at each iteration. In cases in which the trapping time is relatively long, all of these different
methods yielded values which were quite similar. For all average trapping times reported here the width of the
well was taken to be $b=3$~\AA ~and Eq.~(\ref{time1}) shows that $\tau$ scales linearly with $b$.

%%####################################################
%%####################################################
%%####################################################

\section{Approach to Equilibrium}\label{result}

In this section we present a number of calculations that demonstrate the approach towards an equilibrium
distribution of the trapping-desorption fraction.  The parameters are primarily chosen to represent monoenergetic and angularly defined beams of either Ar or
Ne scattering from a tungsten surface.  This leads to a set of guidelines for when one may expect the Maxwell
assumption to be valid, i.e., for when the trapping-desorption fraction  approximates an equilibrium
distribution.

\begin{figure}
\includegraphics[width=6.5in]{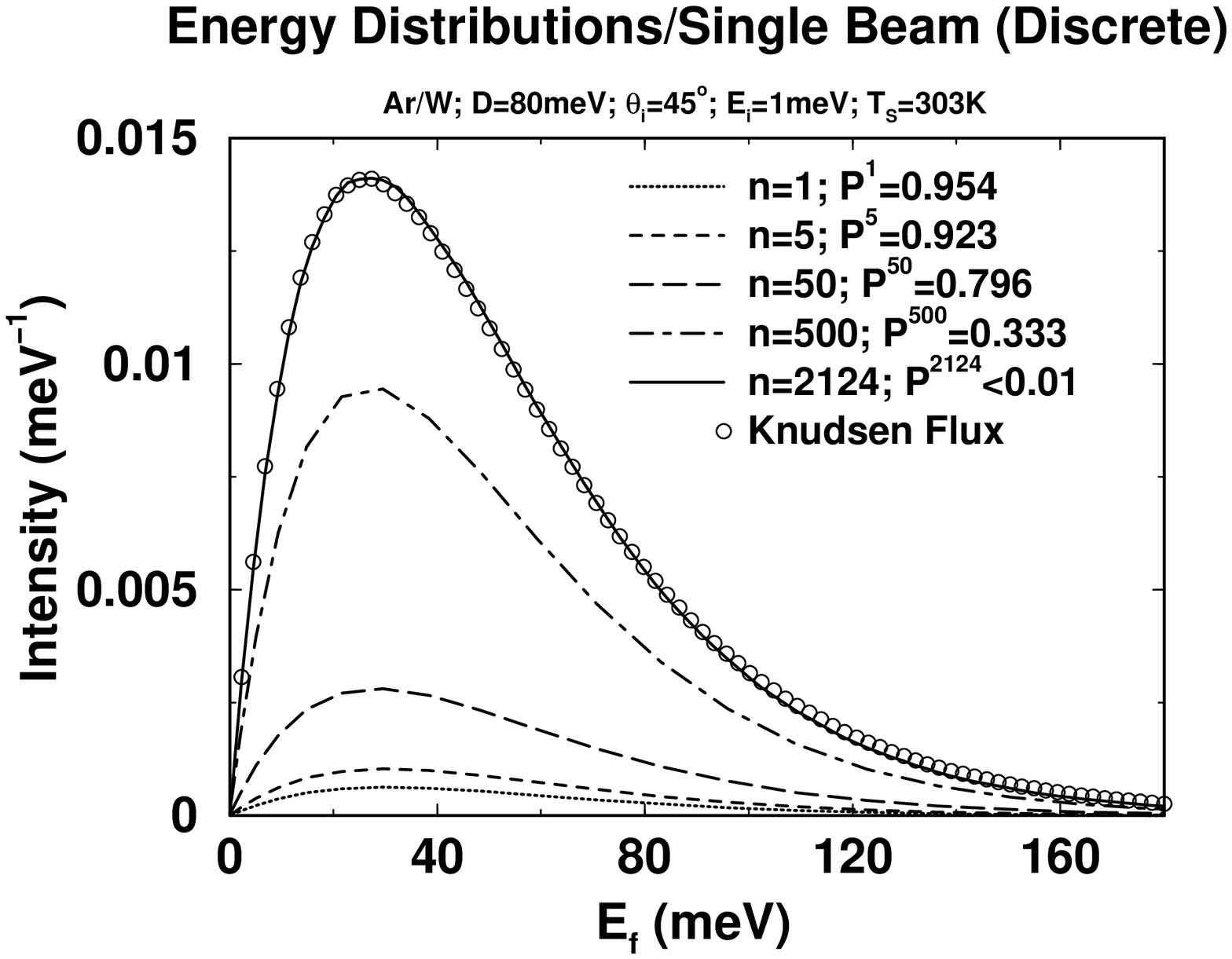}
\caption{Argon scattering from a tungsten surface: the evolution of the final energy distribution for particles
scattered into the continuum states after a specified number of iterations. The surface temperature is $303 $ K,
the incident energy $E_i = 1 $ meV, the well depth $D$ is $80$ meV and the incident angle $\theta _i =
45^\circ$. Five curves of the final distributions for the iteration numbers $N= 1$, $5$, $50$, $500$ and $2124$
are shown. For comparison a Knudsen equilibrium flux is given as open circles. } \label{energy-iter}
\end{figure}

Fig.~\ref{energy-iter} shows an example calculation of the evolution of the energy distribution
as a function of number of iterations
for the case of
argon scattering from a tungsten surface. The incident angle is $45^\circ$, the incident energy is $1 $ meV and
the well depth is chosen to be $80 $ meV and the surface temperature is $303$ K. The dotted curve shows the
continuum energy distribution after the first iteration, which is the second collision with the surface. The
trapping fraction is $P^1= 0.953$, indicating that 95.3\% of the incident particles remain trapped in the
potential well. The dashed and dash-dotted curves show the evolution of the continuum scattered distribution
after increasing numbers of iterations of 5, 50 and 500. After 500 iterations there is still approximately one
third of the incident particles trapped. After 2124 iterations the trapped fraction drops below the arbitrary
threshold of 1\% of the incident particles.

Also shown for comparison in Fig.~\ref{energy-iter} is the Knudsen distribution given by
\begin{equation} \label{Knudsen}
\frac{dK}{dE_f d\Omega_f} =\frac{{E_f \cos \theta_f }}{{\pi (k_B T_G )^2 }} \exp \left\{   - \frac{E_f} {k_B T_S
}   \right\} ~.
\end {equation}
It is clearly seen that the total scattered distribution closely approximates the Knudsen flux when nearly all
the initially trapped particles have been desorbed.

What is actually plotted in Fig.~\ref{energy-iter} is the integral over all final angles of the differential
reflection coefficient, which is essentially the average energy distribution scattered over all outgoing angles.
However, the energy distribution at any fixed final polar and azimuthal angle behaves quite similarly and at the
maximum iteration number the dependence of the distribution on energy is essentially the same at all angles.
This is in agreement with the Knudsen flux which has exactly the same energy dependence at all final angles.

The average trapping time as calculated using the average normal speed from Eq.~(\ref{spav1}) in order to reach the arbitrary threshold of 1\%
of the particles still remaining trapped is $\tau \approx 4.4 \times 10^{-9}$ s.
If the number of iterations is extended to
larger than the $N=2124$ shown in Fig.~\ref{energy-iter} there is essentially no change in the scattered
distribution because the number of particles remaining in the well is insignificant.

For incident energies small compared to the well depth $D$ the final, converged energy distribution after a
large number of iterations is independent of incident energy and incident angle. However, the distributions
calculated after only a small number of iterations will vary somewhat with the choice of these incident
parameters.

\begin{figure}
\includegraphics[width=6.5in]{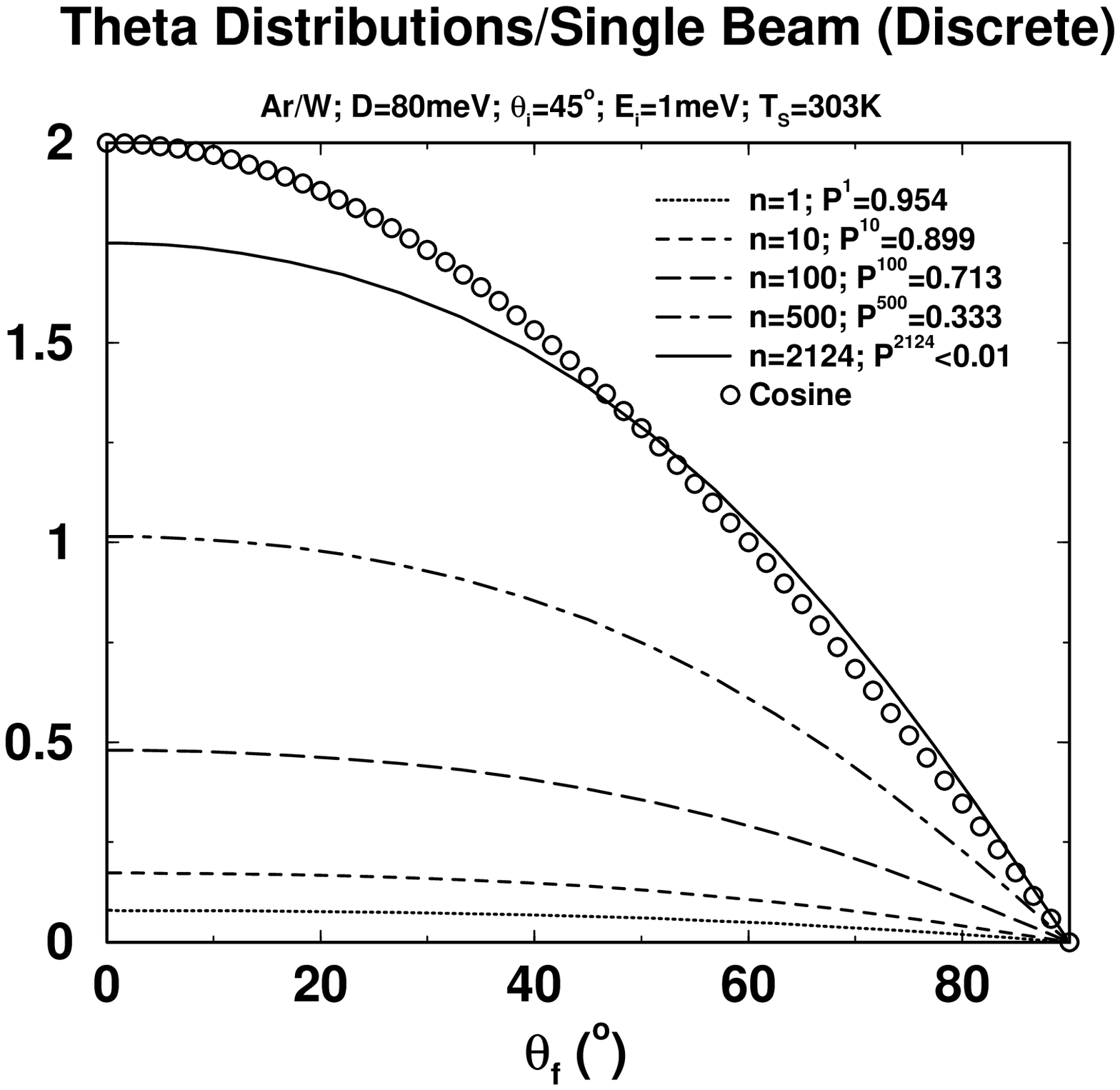}
\caption{ The evolution of the final distribution in polar angle $\theta_f$ for the Ar/W system with the same
parameters as in Fig.~\ref{energy-iter}. A Knudsen cosine distribution is shown as open circles. }
\label{theta-iter}
\end{figure}

The evolution of the angular distribution for the same Ar/W system as shown in Fig.~\ref{energy-iter} is given
in Fig.~\ref{theta-iter}. It is seen that as the number of iterations increases the angular distribution as a
function of the polar angle $\theta_f$ gradually approaches the $\cos \theta_f$ form of the Knudsen flux which
is shown as open circles. The calculations of Figs.~\ref{energy-iter} and~\ref{theta-iter} very quickly become
independent of the azimuthal angle $\phi$ even for very small numbers of iterations. This is consistent with the
behavior of the Knudsen flux which is independent of azimuthal angle.

For the set of initial conditions chosen in Fig.~\ref{theta-iter} it is seen that the  angular distribution does
not achieve a Knudsen distribution even at the largest iteration number calculated.  The shape closely resembles
a cosine function but, for example, its value at $\theta _f = 0$ is only about 88\% of that for the
corresponding Knudsen cosine. This behavior is typical of many of the systems examined, if the well depth is
sufficiently large, and the incident energy small compared to $D$ and $ k_B T_S$ the system will usually achieve
an equilibrium energy distribution after large numbers of iterations and it becomes independent of azimuthal angle
very quickly, but the polar angular distribution is usually very slow to converge to the equilibrium cosine
distribution as seen in Fig.~\ref{theta-iter} and also in Fig.~\ref{arw-theta-0} below.

\begin{figure}
\includegraphics[width=6.5in]{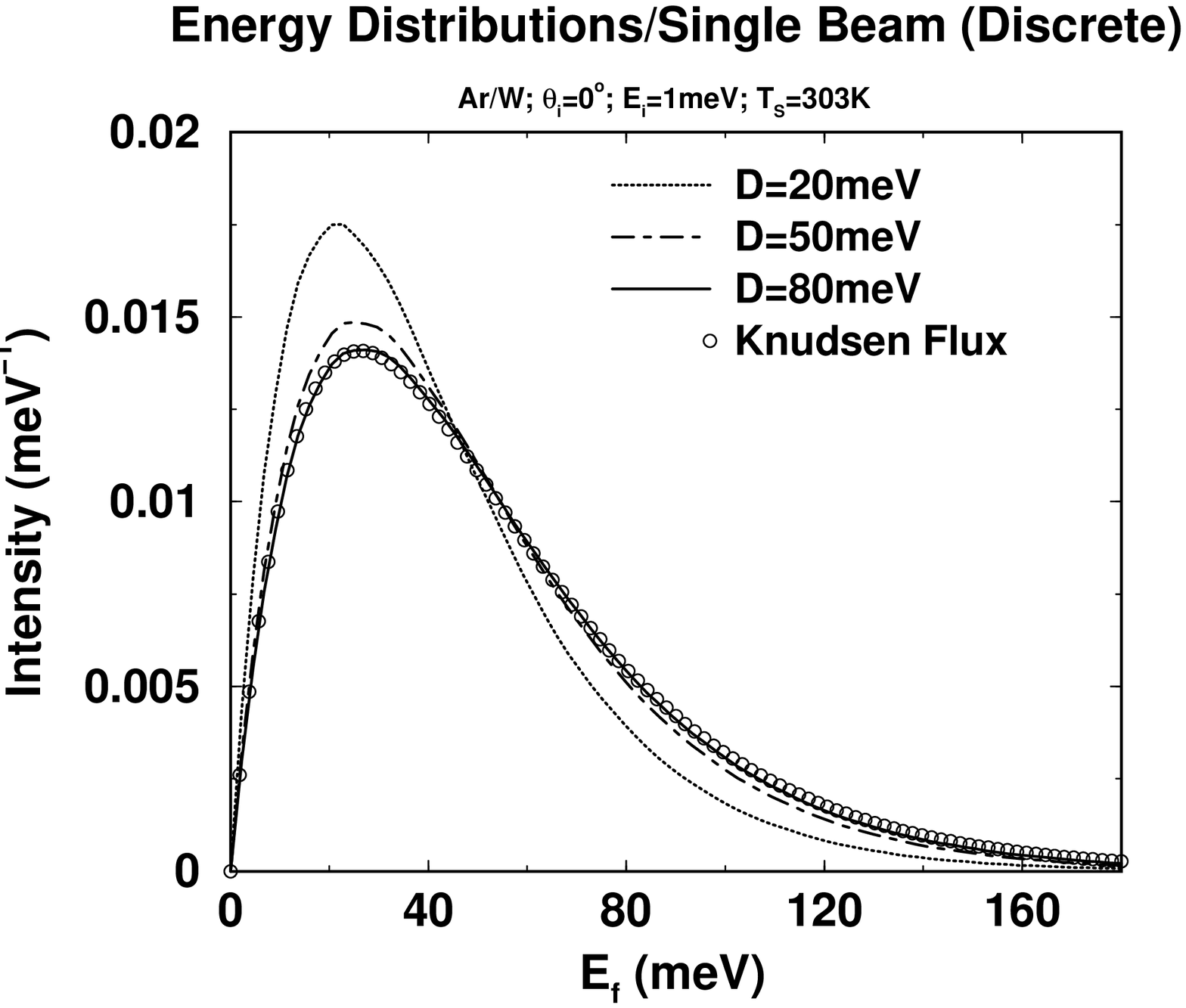}
\caption{Final energy distributions Ar/W system as a function of potential well depths $D = 20 $ meV, $50 $ meV
and $80 $ meV as shown. The temperature of the surface is $303 $ K, the incident energy $E_i = 1 $ meV and
$\theta _i = 0^\circ$. A Knudsen distribution is shown as open circles.
} \label{arw-energy-0}
\end{figure}

The evolution of a system towards a final equilibrium distribution as a function of potential well depth $D$ is
shown in Fig.~\ref{arw-energy-0}. The parameters are similar to the Ar/W system of Fig.~\ref{energy-iter} above,
the projectile is incident normally at $\theta_i=0^\circ$ with $E_i=1$ meV and $T_S=303$ K but completely
converged calculations (meaning less than 1\% of the incident particles remain trapped) are shown for the three
different well depths of 20 meV, 50 meV and 80 meV. This figure shows clearly that even if the incident energy
is very small and the initially trapped fraction is large, the total scattered distribution does not approach
equilibrium unless the adsorption well is sufficiently deep to cause long average trapping times. For a shallow
well of 20 meV with an average  desorption time of about $5.3 \times 10^{-10}$ s
as shown in Table~\ref{arwtime}.
The final distribution deviates strongly from an equilibrium distribution. It is only when the well depth is
increased to about 80 meV, with the corresponding average trapping time of about $4.4 \times 10^{-9}$ s,
that near-equilibrium conditions are achieved.

The response of the angular distribution of the final scattered particles for the same conditions as shown in
Fig.~\ref{arw-energy-0} is given in Fig.~\ref{arw-theta-0}. The progression towards a cosine distribution is
clearly evident with increasing well depth, but even for the quite large value of $D=200$ meV, for which the
average trapping time is $6.4 \times 10^{-7}$ s, the result deviates somewhat from cosine behavior. At
$\theta_f=0^\circ$ the calculated value is 96\% of the cosine maximum.

An example showing the effect of mass on the convergence towards equilibrium is shown in Fig.~\ref{new-energy-0}
corresponding to Ne scattering from tungsten at  1 meV of incident energy. For this system the mass ratio is
roughly half that of Ar/W and approximate equilibrium behavior is not achieved unless the well depth is
approximately 150-200 meV in depth. The corresponding average trapping time for $D=200$ meV is about $3.9 \times 10^{-7}$ s as
seen from Table~\ref{newtime}.

The effects of mass on the approach to equilibrium  is even more dramatically exhibited in Fig.~\ref{mu}  which
shows the scattered energy distribution as a function of the projectile to surface mass ratio $\mu$.  The well
depth is chosen to be 50 meV, the incident energy is 1 meV, $\theta_i=45^\circ$ and the temperature is 303 K.
When the mass ratio is small, for instance $\mu=0.005$ corresponds to hydrogen atoms scattering from tungsten,
the total scattered intensity is far from an equilibrium distribution.  However, for a mass ratio of about
one-half the scattered spectra is very nearly in equilibrium.

Larger temperatures tend to make initial trapping more difficult and lead to more rapid desorption, thus at high temperatures the trapping-desorption fraction will deviate more strongly from an equilibrium distribution.  This is shown in Fig.~\ref{temd} where calculations are shown for the same initial conditions as Fig.~\ref{energy-iter}, i.e., low incident energy and a well depth of 80 meV, but three different surface temperatures of 303, 600 and 1200 K.  For the lowest temperature the scattering is in very good agreement with the corresponding Knudsen equilibrium curve, but begins to deviate quite strongly as the temperature is increased.  As the temperature is increased to the point where the trapping-desorption no longer is equilibrium, the average final energy becomes less than the $2 k_B T_S$ value of the Knudsen distribution.

The method of calculation presented here permits an examination of the energy distribution of the trapped
particles at any average time after the initial collision. An example of this is shown in
Fig.~\ref{efinside-figure} for Ar/W with the same initial conditions as in Fig.~\ref{energy-iter}. Both the
negative energy trapped fraction and the positive energy chattering fraction are exhibited and it is seen that
even after a very few iterations a smooth distribution with a maximum in the negative energy range appears. As
the iteration number increases the most probable energy of all of the trapped particles shifts downward towards
the bottom of the well as the trapped particles continue to lose energy to the surface on average. For very
large numbers of iterations the trapped distribution begins to look exponential-like as a function of energy and
reaches a steady state distribution that retains essentially the same functional form but gradually decreases in
total integrated area as more and more particles are desorbed. Interestingly,  just after the initial
collision the positive energy chattering fraction extends outwards to rather large energies with non-zero amplitude
at positive total energies larger than the magnitude of the well depth.

Some representative calculations of average trapping times $\tau$ are shown in Tables~\ref{arwtime}
and~\ref{newtime} based on the assumption of a square well of width $b=3$~\AA. Table~\ref{arwtime} shows
calculations for the Ar/W system of Fig.~\ref{energy-iter} with 1 meV of incident energy and $T_S=303$ K.
The trapping times for several well depths are calculated in two slightly different ways, first by obtaining the
average speed after each iteration as in Eqs.~(\ref{spav1}) and~(\ref{spav2})  and then also from the root mean
square speed.  The two contributions to the total $\tau$ from the
negative energy trapped fraction and from the positive energy trapped fraction are exhibited separately.
Also shown in Table~\ref{arwtime} are the trapped fractions after the initial collision $P^0$.
Table~\ref{newtime} shows similar information calculated for the Ne/W system of Fig.~\ref{new-energy-0} with the
same incident energy and temperature.

For both systems the results are similar. For shallow well depths the trapping times are very short and the
average time spent in negative and positive energy trapped states is comparable. As the well depth is increased,
trapping times increase dramatically and the average time spent in the positive energy chattering state becomes
negligible compared to the average time in the negative energy bound states. This increase in trapping times is
nearly exponential as a function of well depth as is seen in Fig.~\ref{desorptiontimed} which graphs the numbers
presented in Tables~\ref{arwtime} and~\ref{newtime}. Generally, the trapping times based on the rms speed after
each iteration are somewhat smaller than those based on a calculation of the average speed. For Ar/W with a
physically reasonable well depth of around 100 meV the average trapping time for 99\% of the initially trapped
particles to desorb is approximately $10^{-8}$ s.

All of the above calculations have been for energies relatively small compared to the well depths in order to
illustrate the conditions for which the trapping-desorption fraction approaches an equilibrium distribution.
When the incident energy becomes comparable to or larger than the well depth the nature of the scattered
intensity becomes quite different.  As noticed in an important series of experiments first performed some years
ago the intensity often exhibits a double-peaked structure, with a high energy peak due to direct scattering and
a lower energy peak arising from the trapping-desorption fraction.~\cite{Janda-PRL-79} Fig.~\ref{direct} shows
this for a system corresponding to Ar/W with $\theta_i=45^\circ$, $D=80$ meV, $T_S=303$ K and incident energies
ranging from 100 to 500 meV as marked.  The two contributions are shown in separate panels, with the upper panel
giving the direct scattering and the lower panel the trapping-desorption.  With increasing $E_i$ the total
integrated direct scattering becomes larger and the peak becomes broader, the width roughly increasing
proportionately to $\sqrt{E_i}$.  On the other hand the total trapping-desorption fraction becomes smaller and
the shape of the distribution becomes less and less like that of an equilibrium Knudsen curve.    For large
$E_i$ the trapping desorption intensity develops a long high-energy tail although its peak position always
remains near to the most probable energy of the Knudsen distribution.   The situation in which the differential
reflection coefficient exhibits both distinct direct and a trapping-desorption peaks is discussed more in the
next section in the context of comparisons of the present theoretical model with recent experimental data.

\section{Comparisons with Experiment}\label{sibener}

The calculations exhibited in the above section describe the range of initial conditions that  lead to an
equilibrium distribution in the trapping-desorption fraction but no comparisons with experimental data other
than to a Knudsen distribution were made.  However, in order to be credible, the theoretical approach should be
capable of explaining real experiments.   Demonstrating that ability is the objective of this section.

There is a long history of gas-surface scattering experiments using hyperthermal
atoms~\cite{Janda-PRL-79,atom1,atom2,atom3,atom4,Hurst-JCP-85,atom6,atom7,atom8,atom9,atom10,atom11,atom12,atom13,Morris-JCP-2003}
and molecules~\cite{mol1,mol2,mol3,mol4,mol5,mol6,mol7,mol8,mol9,mol10,mol11,mol12,mol13} as projectiles.  If the projectile gas has mass larger than that of hydrogen  or helium such high
energies imply that the scattering will be classical, which means that many phonons will be transferred in the
collision.  This is the type of experiment that should be amenable to the theoretical treatment described here.
In many cases the energy-resolved scattered spectra exhibit a double peaked structure, with a somewhat narrow
high-energy peak centered at smaller energy than the incident beam energy (if the incident energy is large
compared to the surface temperature) and a broader low-energy peak at thermal energies.  The usual
interpretation has been that the high-energy peak is direct scattering from a single collision (or at most, a
very small number of collisions) and the low-energy peak arises from trapping in the physisorption well of the
interaction potential with subsequent desorption at a sufficiently later time so that those particles come into
near equilibrium at the surface temperature.~\cite{Janda-PRL-79}

A recent and important paper reporting extensive measurements that show  clearly a set of conditions for which
direct scattering and trapping-desorption can be observed is that of Gibson, Isa and Sibener for scattering of
Ar from an ordered 1-decanethiol self-assembled overlayer on a Au(111) substrate.~\cite{Sibener-JCP-03}  The
experiments were carried out with well-defined monoenergetic beams of Ar incident at energies ranging from
roughly 60 to 600 meV, and with both incident and detector angles independently variable and ranging from
near-normal to near-grazing with respect to the surface.
All measurements were made in the scattering plane (the plane containing the surface normal and the incident beam) which was aligned along the $ \langle 1  \overline{1}  0   \rangle$ direction of the Au(111) surface.
At low incident energies and if $\theta_i$ or
$\theta_f$ was near-normal they did not observe a clear double peaked intensity in the scattered spectra.
However, at higher energies and for large incident or final angles the characteristic double-peaked structure
was very apparent.  They analyzed their data quantitatively with an ad hoc model consisting of the sum of a
shifted Maxwell-Boltzmann distribution to fit the direct scattering and an equilibrium distribution to fit the
trapping-desorption fraction.  They also made some more qualitative analysis of their data using classical
trajectory calculations developed by Hase and coworkers.~\cite{Hase-96}  In the process of their analysis they
determined, by assuming that the direct scattering was due to a single collision and using well-known Baule
relations for binary collisions, that the effective mass of the surface implied a mass ratio $\mu=0.62$, or
$M_C=64.4$ amu as opposed to the total mass of the 1-decanethiol which is $174.3$ amu.
The potential energy landscape function developed for the classical trajectory calculations had a well depth ranging from 33 meV at the on-top sites above the terminal CH$_3$ groups to 67 meV in the center of the rhombus formed by a group of four of the methyl groups.~\cite{Sibener-JCP-03}

Fig.~\ref{ei653} shows an example of calculations compared to the Ar scattering data taken from the upper panel
of Fig.~2 in Ref.~[\cite{Sibener-JCP-03}] and is for their lowest incident energy $E_i=65.3$ meV.  The data was
reported as intensity versus time-of-flight (TOF) and the calculations have been transformed accordingly.  The
other incident parameters are $\theta_i = 45^\circ$, $\theta_f = 50^\circ$ and $T_S=135$ K.  The calculations
were carried out for a well depth $D=35$ meV and an effective surface mass $M_C=71$ amu  ($\mu = 0.56$).

This effective surface mass ratio is slightly smaller than the value
$\mu = 0.62$ estimated in Ref.~[\cite{Sibener-JCP-03}].   For this case of low incident energy the calculations are not particularly sensitive to the value of $\mu$ because there is very little evidence for a significant direct scattering component.
However the value of $\mu = 0.56$ is chosen  as a consequence of comparisons with the higher energy data shown below, where the much more pronounced direct scattering component is extremely sensitive to $\mu$.
In this
example the data do not exhibit a double-peaked structure and the calculated most probable energy  (peak position) has a
TOF time corresponding to $E_f= 12.57$ meV, very close to the equilibrium value of 11.6 meV at this temperature, again indicating that the scattering is mostly trapping-desorption.

Three calculated curves are shown in Fig.~\ref{ei653}, the solid curve is the total differential reflection coefficient converted to TOF, the dashed curve is the trapping-desorption contribution only, and for comparison a Knudsen equilibrium distribution is included as a dash-dot curve.  The Knudsen distribution and the calculation are both normalized to unit intensity as in Eq.~(\ref{unitarity}). Although it is apparent that the Knudsen curve, if renormalized to fit as closely as possible to the data, would match essentially as well as the total calculated intensity, the fact that it is smaller
and nearly the same as the calculated trapping-desorption fraction
indicates that there is significant direct scattering but its most probable energy and width is nearly the same as the equilibrium distribution.
Because of the strong overlap of the direct and trapping-desorption fractions, it is not surprising that the data of Fig.~\ref{ei653} can be matched roughly as well by a total scattering intensity using a range of well depths from 20 to somewhat over 35 meV.  We have chosen $D=35$ meV because of the much stronger constraints placed on this parameter by the higher energy data considered below.

Three examples of data measured at the intermediate energy of 365 meV are shown in Fig.~\ref{ei365} at the same temperature of 135 K and for three different combinations of incident and final angles.
The middle panel for $\theta_f = 50^\circ$ and $\theta_i = 30^\circ$, relatively close to normal incidence,   does not exhibit a double peaked structure.  The other two panels, for $\theta_i = 45^\circ$ and
$\theta_f = 50^\circ$ (upper panel) and  $\theta_i = 30^\circ$ and
$\theta_f = 80^\circ$ (lower panel) present a clear distinction between the rather sharp peak at short TOF and a broader shoulder at larger times.
These data were taken from the middle panel of Fig.~2 and the lower two panels of Fig.~4, respectively, of Ref.~[\cite{Sibener-JCP-03}].
The solid curves in Fig.~\ref{ei365} are calculations carried out with $\mu = 0.56$ and $D=35$ meV.
The calculations explain the data quite well, and  they show clearly the separation between the direct and trapping-desorption fractions.    The value $D=35$ meV agrees well with the that of the potential energy function for this system developed in Ref.~[\cite{Sibener-JCP-03}]

 Also shown in Fig.~\ref{ei365} are the trapping-desorption fraction and the Knudsen curves.  Interestingly, the trapping-desorption fraction itself has a multiple-peaked structure with a small sub-peak appearing at almost the same final energy as the direct scattering contribution.  This small high-energy sub-peak comes from the first few collisions as the initially adsorbed particles travel in the potential well.  These first few collisions have a high probability of ejecting particles back into the continuum with relatively little loss of energy compared to the direct scattering fraction.  However, it is clear that the largest part of the trapping-desorption fraction resembles closely  the shape of the Knudsen curve.

It also becomes apparent from Fig.~\ref{ei365}  that there is a straightforward manner in which the comparison of calculations with data allows for the determination of the two parameters.  The effective mass determines the most probable final energy of the direct contribution and then the well depth determines the relative intensity of the trapping-desorption fraction which becomes bigger with increasing $D$.
The peak position of the direct scattering contribution is extremely sensitive to the mass ratio and this is why we chose the value $\mu = 0.56$ as opposed to the value $\mu = 0.62$ of Ref.~[\cite{Sibener-JCP-03}] which was based on the Baule equations describing hard sphere scattering.

Three examples of data for scattering at the high energy $E_i=582$ meV, all of which exhibit the double-peaked structure, are shown in Fig.~\ref{ei582}.
The data were taken from the lower panel of Fig.~2 and the middle and lower panel of Fig.~3, respectively, of Ref.~[\cite{Sibener-JCP-03}].
The upper panel of Fig.~\ref{ei582} is for $\theta_i=45^\circ$ and $\theta_f=50^\circ$, the middle panel is for $\theta_i=45^\circ$ and $\theta_f=40^\circ$, and the lower panel is for $\theta_i=60^\circ$ and $\theta_f=40^\circ$.

In the upper two panels with $\theta_i=45^\circ$ two curves showing the total scattering intensity are shown, for $D=35$ and 45 meV while in the lower panel with $\theta_i=60^\circ$ only tht $D=35$ meV calculation is shown.  All calculations were done with $\mu=0.56$, the value that leads to agreement with the data for the direct scattering peak.  The dotted curve is the Knudsen distribution and the dash-dotted curves show the trapping-desorption fraction.

It is interesting that at this larger incident energy the two cases with the more normal incident angle of $45^\circ$ require a well depth of 45 meV in order to obtain agreement between calculations and data, while for the much more grazing incidence of $80^\circ$ the best well depth is 35 meV, the same as used for all the lower energy calculations.  This appears to indicate that for larger normal incident energy the incoming atoms are probing deeper parts of the potential energy landscape, which as mentioned above has been estimated to have a well with depths that vary between 33 and 67 meV.~\cite{Sibener-JCP-03}
Again, as in  Fig.~\ref{ei365} the trapping-desorption fraction exhibits structure at larger final energies near the energy of the direct scattering.

Comparison of the present calculations to this Ar scattering data leads to a few general comments that can be applied to the observed energy-resolved spectra for cases in which  a double structure appears due to the distinct phenomena of direct scattering and trapping-desorption: \\
(1) A characteristic double-feature structure, with a well-defined direct scattering peak and a secondary peak or shoulder arising from trapping-desorption appears only at relatively high incident beam energy and when one or the other of $\theta_i$ or $\theta_f$ is large, as was already clear from Ref.~[\cite{Sibener-JCP-03}].
It is also necessary that the physisorption well depth is sufficiently large to cause significant trapping in the bound states during the initial collision, and the temperature must be smaller than $D$.   However, for  near normal incidence conditions and with a large well, the initial trapping becomes so large that the direct scattering contribution becomes small, and this explains the need for the incident angle to be relatively large.  This situation becomes evident in Fig.~\ref{thf10} which shows calculations for $E_i = 582$ meV with $\theta_i=\theta_f=10^\circ$ and $D=35$ meV.  Even though the incident energy is large compared to the well depth, the normal incidence conditions gives rise to such large trapping that the total scattering is not very different from the trapping-desorption fraction. \\
(2) When the direct scattering contribution is significant, the trapping-desorption intensity deviates substantially from that of an equilibrium Knudsen distribution.  In fact, the trapping-desorption signal can exhibit structure and small peak-like features at high energies close to those of the direct contribution as shown in Figs.~\ref{ei365} and~\ref{ei582}.  However, in many cases the largest portion of the  trapping-desorption intensity resembles the shape of a Knudsen distribution and this is especially true for the low energy tail. \\
(3) It is interesting to note that the direct and trapping-desorption fractions should have very characteristic
and  quite different signature behaviors in their temperature dependence.  The direct scattering, which in the
present calculations arises from a single collision with the surface, is essentially given by Eq.~(\ref{dis}).
Under conditions for which the direct scattering of Eq.~(\ref{dis}) appears nearly Gaussian-like in the energy
transfer, which is the situation in several of the cases shown here, the mean square energy deviation which is
proportional to the full width at half-maximum (FWHM) is~\cite{dis1,Manson-Muis,Burke-Kohn}
\begin{eqnarray} \label{M4a}
 \langle E_f^2 \rangle = \frac{FWHM}{8 \ln(2)}  = 2 g(\mu, \theta) E_i k_B T_S
 ~,
 \end {eqnarray}
where
\begin{eqnarray} \label{M4}
g(\mu, \theta) ~=~ \frac{ \mu\left( 1 + f(\mu, \theta) - 2 \sqrt{f(\mu, \theta)} \cos \theta \right)  }
{\left( 1 + \mu - \frac{\mu \cos \theta}{\sqrt{f(\mu, \theta)}}  \right)^2}
~,
\end {eqnarray}
with
\begin{eqnarray} \label{M3}
f(\mu, \theta) ~=~ \left( \frac{\sqrt{1-\mu^2 \sin^2\theta} + \mu \cos \theta}{1+ \mu}  \right)^2
~,
\end {eqnarray}
where $\theta$ is the total scattering angle (i.e., the angle between ${\bf p}_i$ and ${\bf p}_f$) and Eqs.~(\ref{M4a})-(\ref{M3}) are obtained under the assumption of binary collision conditions
for which $\overline{E_f} = f(\mu, \theta)  E_i $.
Thus, Eq.~(\ref{M4a}) shows that the FWHM of the direct scattering peak will be approximately proportional to $\sqrt{T_S}$, which is the characteristic of the multiphonon scattering regime.  However, the trapping-desorption fraction will have a FWHM temperature dependence more closely approximating the linear in $T_S$ behavior of the Knudsen distribution of Eq.~(\ref{Knudsen}).  Similarly, the most probable intensity (maximum peak intensity) of the direct scattering will vary as $1/\sqrt{T_S}$ according to Eq.~(\ref{dis}) while the trapping-desorption peak intensity should behave more like that of the $1/T_S$ behavior of the Knudsen distribution of Eq.~(\ref{Knudsen}). \\
(4) Finally, the comparison of the present calculations with the data provide a simple way to extract the two relevant parameters.  The position in final energy of the narrow direct scattering peak determines the value of the effective surface mass, and in fact, the most probable energy is quite sensitive to this parameter.
The intensity of the broad trapping-desorption peak increases with the well depth, and fixing the relative intensities of the two contributions determines $D$.  This indicates that for the large incident energies considered here the primary influence of the well depth is to establish the initial trapping fraction.  Once trapped, the details of the shape of the potential well are not important as is evidenced by the fact that the long-time tail trapping-desorption fraction (the low energy tail) eventually desorbs at thermal energies.

\section{Conclusions}\label{conclusion}

In this paper we have developed a theoretical formulation of the scattering of atomic projectiles with surfaces
that includes not only the direct scattering arising from a single, or a small number of collisions with the
surface but also allows for trapping and subsequent collisions of trapped particles inside the physisorption well.
The trapped particles can be followed until they eventually all desorb and leave the surface region.  The
multiple collisions of the initially trapped fraction with the surface are treated with an iteration algorithm
that tracks trapped particles with both negative and positive total energies and determines at each subsequent
collision the fraction scattered back into the positive energy continuum which then leaves the surface region.

Using this theoretical formalism we have calculated numerous examples, firstly in order to establish the
conditions  under which the trapped and subsequently desorbed particles approach an equilibrium distribution,
i.e., to establish the conditions under which the Maxwell assumption is valid, and secondly we have used the
theory to produce quantitative agreement with recent measurements thus providing  explanations of the basic
underlying processes that give rise to the experimental scattered spectra.

Under many conditions, the observed spectra in gas-surface scattering experiments consist  of two distinct
contributions. The first contribution is the direct scattering part which is usually a relatively sharp peak with
a most probable energy somewhat lower than the incident energy provided that the surface temperature is not
large compared to the incident energy.  The second of these contributions is the trapping-desorption, attributed
to particles that are initially trapped and then spend a large time moving in the physisorption well where they
slowly begin to exchange energy with the surface and then eventually desorb in a distribution at thermal
energies corresponding to the surface temperature.  A large part of the work considered here is devoted to
determining when the trapping-desorption fraction approaches an equilibrium Knudsen flux.

Our calculations show that under conditions in which a clear direct scattering and trapping-desorption
double-peaked structure is evident in the energy-resolved spectra the trapping-desorption fraction,
although mainly emitted at thermal energies,
 can differ
considerably from an equilibrium distribution.  It can even exhibit structure consisting of small peaks at high
energy near the most probable energy of the direct scattered intensity.  These higher energy  peaks arise from the first few
collisions with the surface inside the well indicating that these initial collisions have a large probability of ejecting
particles into the continuum with little additional energy loss as compared to the direct scattering.  However,
even under conditions for which the trapping-desorption fraction is highly non-equilibrium its low energy tail
still is well described by a Knudsen distribution. Thus, our calculations show that the Maxwell assumption is
rarely achieved in real experimental conditions, although it is very useful as an approximate guide as is
evident from the fact that it is still often used as a method to analyze measured data.

We carried out a number of calculations in order to characterize the conditions under which  the
trapping-desorption fraction does approach an equilibrium distribution.  Basically, equilibrium behavior is
achieved only for cases where the direct scattering is negligible and average trapping times are long, which implies that nearly all of the incident
beam is adsorbed after the first collision.  This implies an incident energy relatively small compared to the
interaction potential physisorption well depth and temperatures corresponding to energies (measured in units of
$k_B T_S$) that are also small compared to the well depth.  The approach to equilibrium occurs more rapidly with
larger gas-to-surface-atom mass ratios when this ratio is less than unity.

The approach to equilibrium of the trapping-desorption fraction was studied as a function of of all  the initial
experimental parameters that can be manipulated, including the projectile and surface mass, the well depth of
the potential, the incident energy and angles, the final scattering angles and the surface temperature.  For
example, as the well depth is increased with all other parameters held constant, we find that the
energy-resolved scattered spectrum rather quickly approaches that of a Knudsen distribution when the well depth
becomes significantly larger than the incident energy provided the temperature is also small compared to the
well depth.  The angular behavior becomes independent of azimuthal angle under the same conditions that the
energy distribution becomes Knudsen-like.  However, the polar angle $\cos \theta_f$  behavior is only
approximately obeyed for conditions under which the energy dependence first approaches equilibrium, even though
the energy dependence is nearly Knudsen-like at all polar angles.  Only for well depths very large compared to
the incident energy and temperature does the trapping-desorption fraction achieve the classic Knudsen $\cos
\theta_f$ shape.

In this formalism, because of its iterative approach, the number of collisions of the trapped  particles with the
surface can be followed.  This means that trapping times can be calculated as well as other information such
as the relative fractions of particles trapped with negative total energies and those trapped in the chattering
states having  positive total energies.   Under conditions in which the trapping-desorption fraction did
achieve near equilibrium the trapping times were estimated to be as large as $10^{-8}-10^{-7}$ s.  The positive
total energy chattering fraction can be large under conditions where equilibrium is not achieved, but for
conditions that produce equilibrium in the trapping-desorption the fraction of trapped particles residing in the
chattering states becomes negligible.

An important aspect of this work is that the theoretical model used for the calculations provides quantitative
explanations of experimental measurements. We made comparisons with   important and recent Ar scattering data
obtained in beams-surface scattering experiments with a surface consisting of a self-assembled adsorbed layer of
1-decanethiol on a well-ordered Au(111) substrate.~\cite{Sibener-JCP-03} In agreement with the experimental
observations, we found that clearly distinguishable direct and trapping-desorption contributions  arise when the
incident beam energy is large compared to the potential well depth and when one or the other of the incident or
detector angles was large relative to the surface normal.  In addition the calculations indicate that, in order
to resolve distinct direct and trapping-desorption features,  the temperature must be small compared to the well
depth and the well depth must be large enough to cause trapping of a significant fraction of the incident beam
at the initial collision.

Under conditions for which distinct direct and trapping-desorption features were  evident, the data can be used
to determine two important characteristics of the interaction potential, the effective surface mass of the
adsorbate and the well depth.  The effective mass is determined by matching the calculated direct scattering
peak to that of the experiment, and the well depth then is determined by matching the relative intensity of the
trapping-desorption contribution.

An interesting prediction coming out of this work is that the direct and trapping-desorption  contributions have
clearly different signature behaviors as a function of surface temperature.  The FWHM of the direct peak should
increase approximately with the square root of the temperature whereas the trapping-desorption has a full width
that increases approximately linearly with $T_S$, similarly to the Knudsen distribution. The most probable
intensity of the direct peak, according to the scattering model used here, decreases inversely with the square
root of temperature, while the trapping-desorption decreases roughly linearly with the inverse of the
temperature, again similarly to the Knudsen distribution.  For both peaks, the increase in FWHM and decrease of
most probable intensity is the behavior expected in order to preserve unitarity.  It should be noted that the
direct scattering, as shown in the approximations to Eq.~(\ref{dis}) appearing in Eqs.~(\ref{M4a})-(\ref{M3}),
exhibits the same behavior in the incident energy $E_i$  as it does in the temperature
$T_S$.  However, the most appropriate parameter in which to carry out an experimental search for these behaviors would be the temperature, because the interaction potential surface is likely to change  with variation of
$E_i$ but is less likely to change with $T_S$.

This work demonstrates that calculations of direct scattering, trapping and desorption  in atom-surface scattering can
provide real quantitative explanations of experiments as well as indicating the conditions for the validity of
Maxwell's assumption on the equilibrium nature of the trapping-desorption fraction.  It shows that the
interaction potential model must contain two essential ingredients:  a physisorption well depth and  it must allow for
transfer of mechanical energy between the projectile and the surface atoms.  However, it also shows that
the most important aspect is to have a theory in which the statistical mechanics is treated in a reasonably
correct manner.

%XXXXXXXXXXXXXXXXXXXXXXXXXXXXXXXXXXXXXXXXXXXXXXXXXXXXXXXXXXXXXXXXXXXXXXXXXXXXXXXXXXXX

%%XXXXXXXXXXXXXXXXXXXXXXXXXXXXXXXXXXXXXXXXXXXXXXXXXX
\begin{table}[htbp]
\centering \caption{
The desorption times $\tau$ and initial sticking fractions $P^0$ for Ar/W. $E_i$ is 1 meV, $\theta _i$ is $45^\circ$, $T_S$ is 303K and well depths ranging from 20 meV to 200 meV are shown.  The upper set of values was obtained from the normal rms speed of trapped particles and the lower set are values determined from the average normal speed.  }
\begin{tabular} {|c|c|c|c|c|}\hline
Ar/W & $\tau _{rmsC}$ & $\tau _{rmsT}$ & $\tau_{rms} $ & $P^0$ \\
\hline
D=20meV & 1.28e-10 & 1.66e-10 & 2.94e-10 & 0.825 \\
\hline
D=50meV & 1.26e-10 & 6.63e-10 & 7.89e-10 & 0.933 \\
\hline
D=70meV & 1.35e-10 & 1.58e-9 & 1.72e-9 & 0.958 \\
\hline
D=80meV & 1.40e-10 & 2.42e-9 & 2.56e-9 & 0.966 \\
\hline
D=200meV & 1.99e-10 & 3.59e-7 & 3.59e-7 & 0.995 \\
\hline
$\tau$ ($\bar v$) & $\tau _{C}$ & $\tau _T$ & $\tau $ & ~ \\
\hline
D=20meV & 2.27e-10   &  3.04e-10 &  5.31e-10  &  ~ \\
\hline
D=50meV & 2.10e-10   &  1.16e-9 &  1.37e-9   & ~ \\
\hline
D=70meV & 2.21e-10     &  2.74e-9 &  2.96e-9 & ~  \\
\hline
D=80meV & 2.29e-10  &  4.20e-9 &  4.43e-9 & ~ \\
\hline
D=200meV & 3.26e-10  &  6.40e-7 &  6.40e-7  & ~ \\
\hline
\end{tabular}\label{arwtime}
\end{table}
%%XXXXXXXXXXXXXXXXXXXXXXXXXXXXXXXXXXXXXXXXXXXXXXXXXX

\begin{table}[htbp]
\centering
\caption{
The desorption time for Ne/W with other parameters the same as in Table~\ref{arwtime}. }
\begin{tabular} {|c|c|c|c|c|}\hline
Ne/W & $\tau _{C}$ & $\tau _T$ & $\tau $ & $P^0$ \\
\hline
D=20meV & 9.56e-11 & 1.35e-10 & 2.30e-10 & 0.848 \\
\hline
D=30meV & 9.50e-11 & 2.10e-10 & 3.05e-10 & 0.891 \\
\hline
D=70meV & 9.47e-11 & 1.11e-9 & 1.20e-9 & 0.954 \\
\hline
D=80meV & 9.70e-11 & 1.66e-9 & 1.76e-9 & 0.961 \\
\hline
D=150meV & 1.16e-10 & 2.88e-8 & 2.89e-8 & 0.984 \\
\hline
D=200meV & 1.29e-10 & 2.22e-7 & 2.22e-7 & 0.990\\
\hline
$\tau$ ($\bar v$)& $\tau _{C}$ & $\tau _T$ & $\tau $ &  \\
\hline
D=20meV & 1.73e-10    &  2.42e-10 &  4.16e-10 &  \\
\hline
D=30meV & 1.59e-10   &  3.70e-10 &  5.29e-10 &   \\
\hline
D=70meV & 1.58e-10   &  1.92e-9 &  2.07e-9 &   \\
\hline
D=80meV & 1.61e-10   &  2.88e-9 &  3.04e-9  & \\
\hline
D=150meV & 1.89e-10   &  5.07e-8 &  5.08e-8 &   \\
\hline
D=200meV & 2.10e-10  &  3.93e-7 &  3.93e-7  & \\
\hline
\end{tabular}\label{newtime}
\end{table}

\begin{figure}
\includegraphics[width=6.5in]{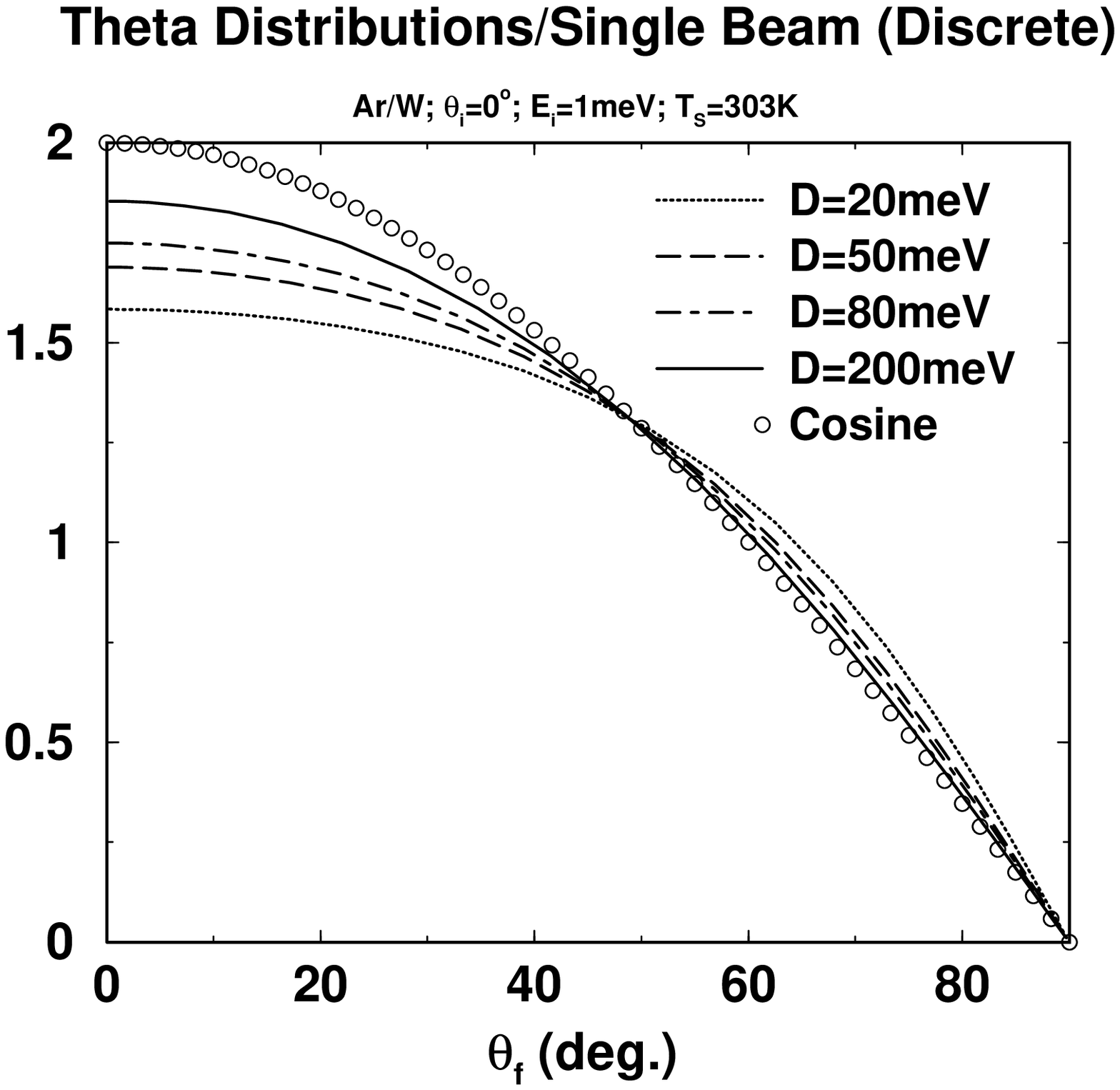}
\caption{The polar angular distribution for the same system shown in Fig.~\ref{arw-energy-0}. The evolution
towards the Knudsen distribution, displayed as open circles, is shown for a series of increasing well depths.}
\label{arw-theta-0}
\end{figure}

\begin{figure}
\includegraphics[width=6.5in]{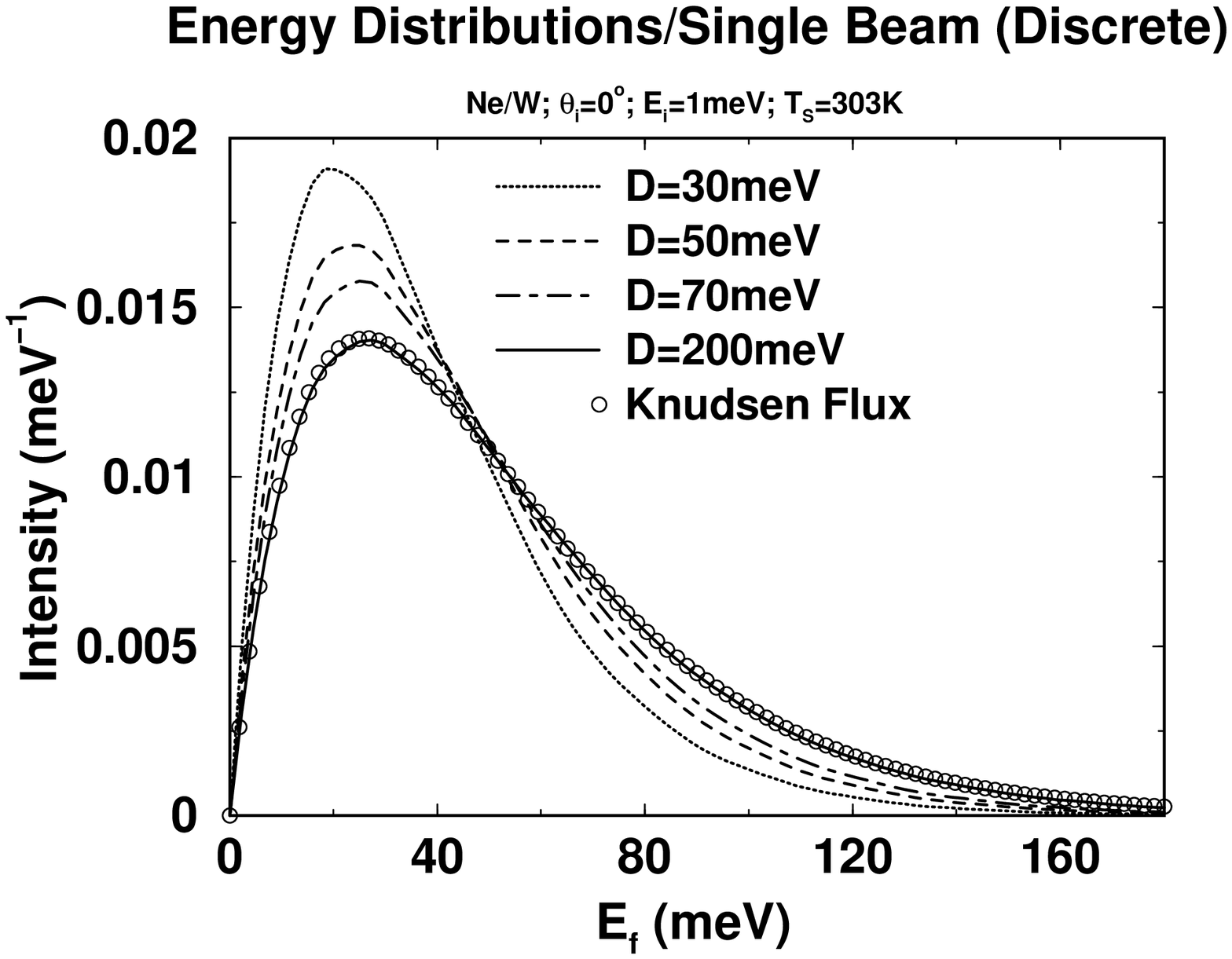}
\caption{Energy distribution of the scattered particle for Ne/W for several well depths as shown. The
temperature of the surface is $303 $ K, the incident energy $E_i = 1 $ meV and $\theta _i = 0^\circ$. The
Knudsen distribution is shown as open circles. } \label{new-energy-0}
\end{figure}

\begin{figure}
\includegraphics[width=6.5in]{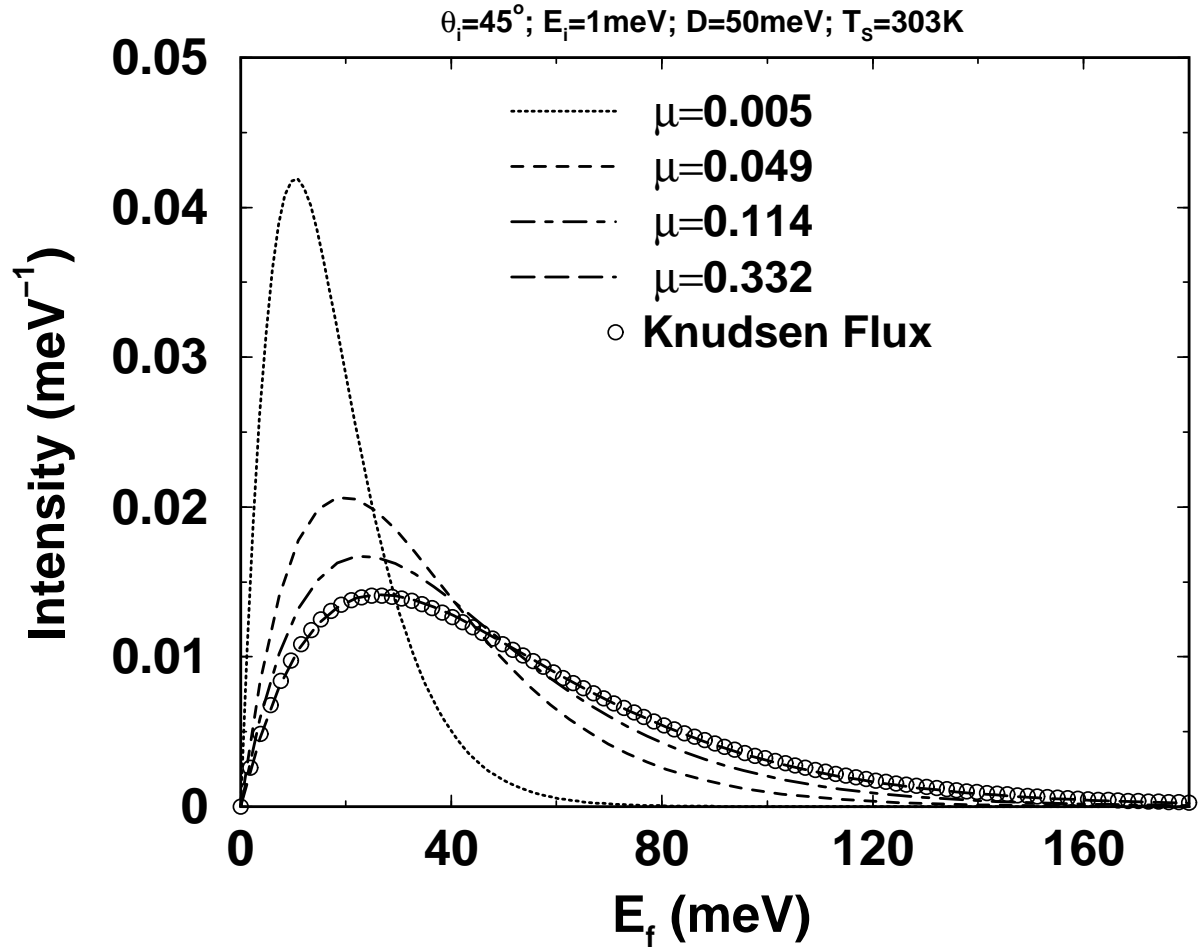}
\caption{The energy distribution  as a function of the mass ratio $\mu$ for a system with $\theta_i = 45^\circ$,
$E_i=1$ meV, $T_S=303$ K and $D=50$ meV.
} \label{mu}
\end{figure}

\begin{figure}
\includegraphics[width=5.0in]{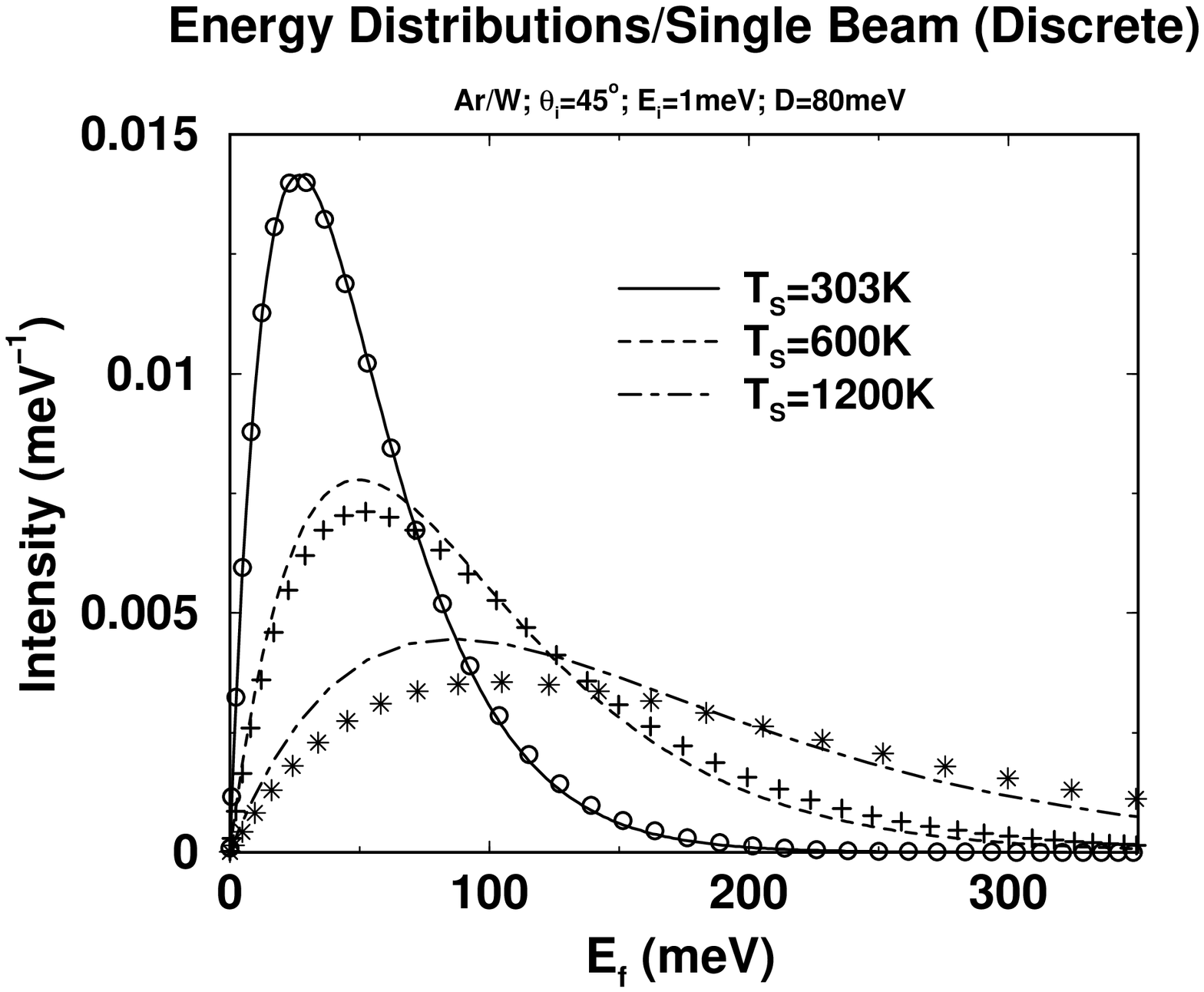}
\caption{Energy distribution of the scattered particles for Ar/W
with incident energy $E_i = 1 $ meV, $\theta _i = 45^\circ$ and a well depth $D= $ 80meV.
Three different surface temperatures $T_S = $ 303 K, 600 K and 1200 K are shown as curves and the corresponding equilibrium distributions are shown as points.
}
\label{temd}
\end{figure}

\begin{figure}
\includegraphics[width=6.5in]{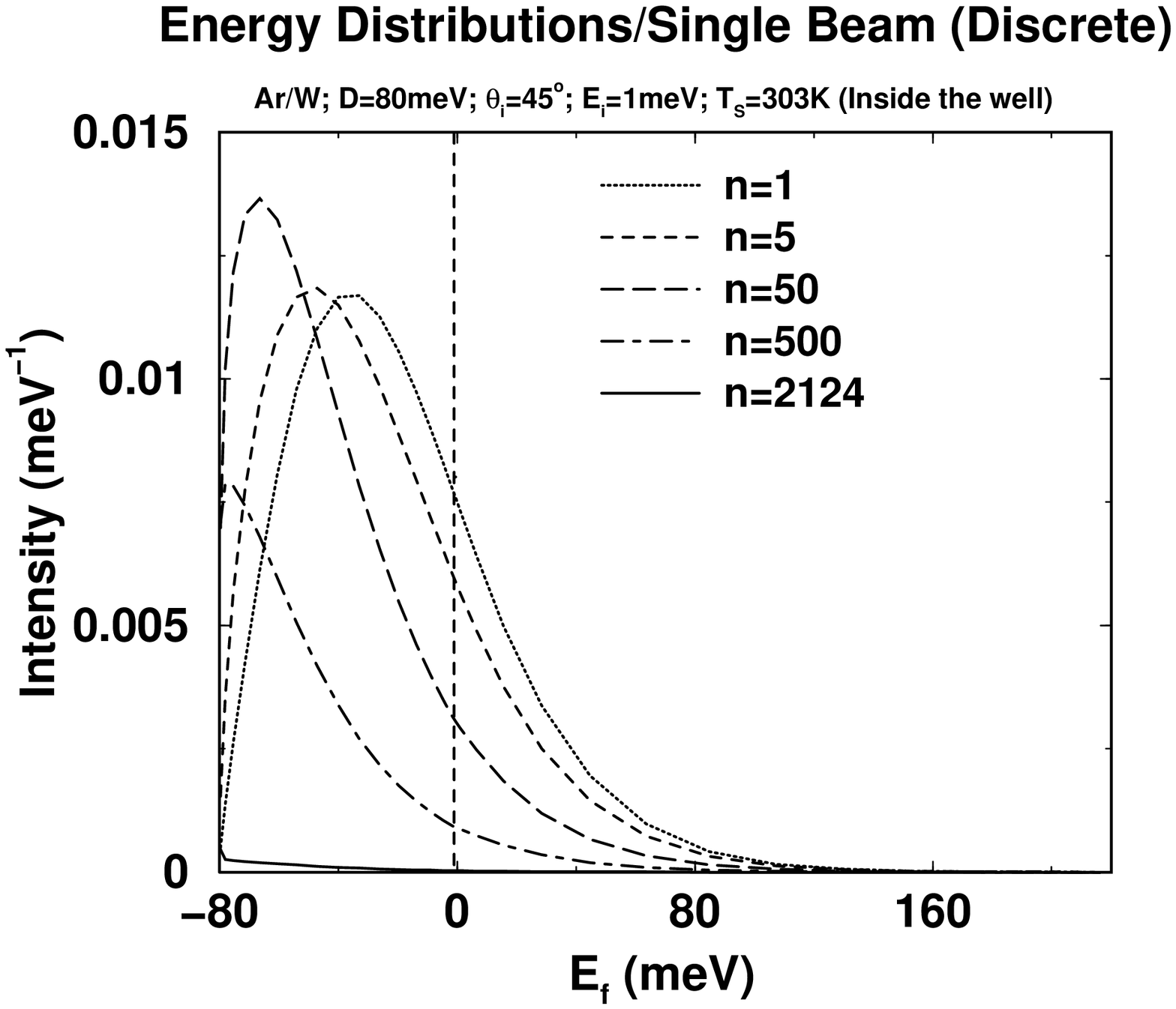}
\caption{ The evolution of the energy distribution inside the potential well as a function of iteration number
for the same system as in Fig.~\ref{energy-iter}.
} \label{efinside-figure}
\end{figure}

\begin{figure}
\includegraphics[width=6.5in]{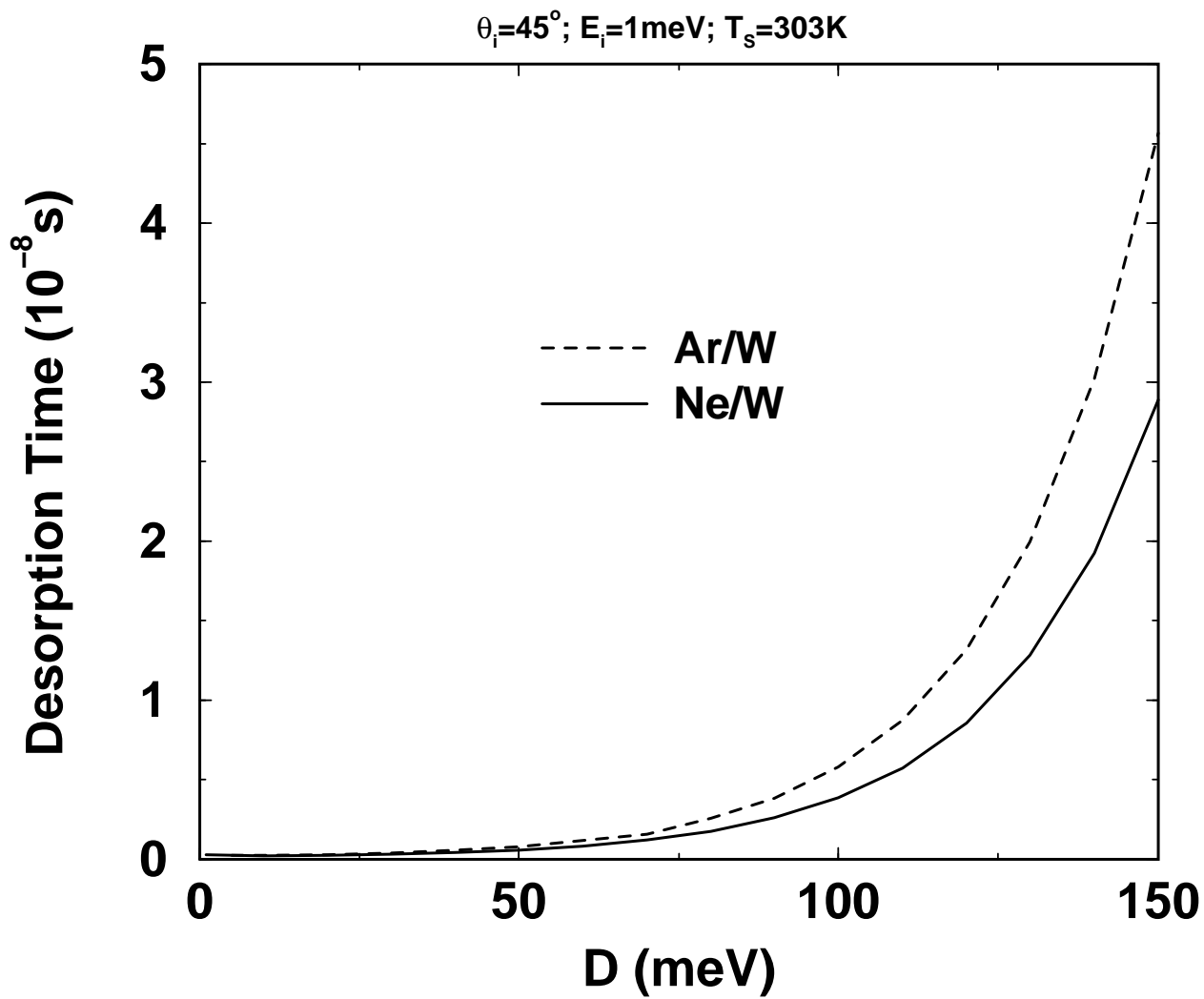}
\caption{The desorption times for Ar/W and Ne/W as  functions of the well depth for the incident conditions shown in Tables~\ref{arwtime}
and~\ref{newtime}.
} \label{desorptiontimed}
\end{figure}

\begin{figure}
\includegraphics[width=6.5in]{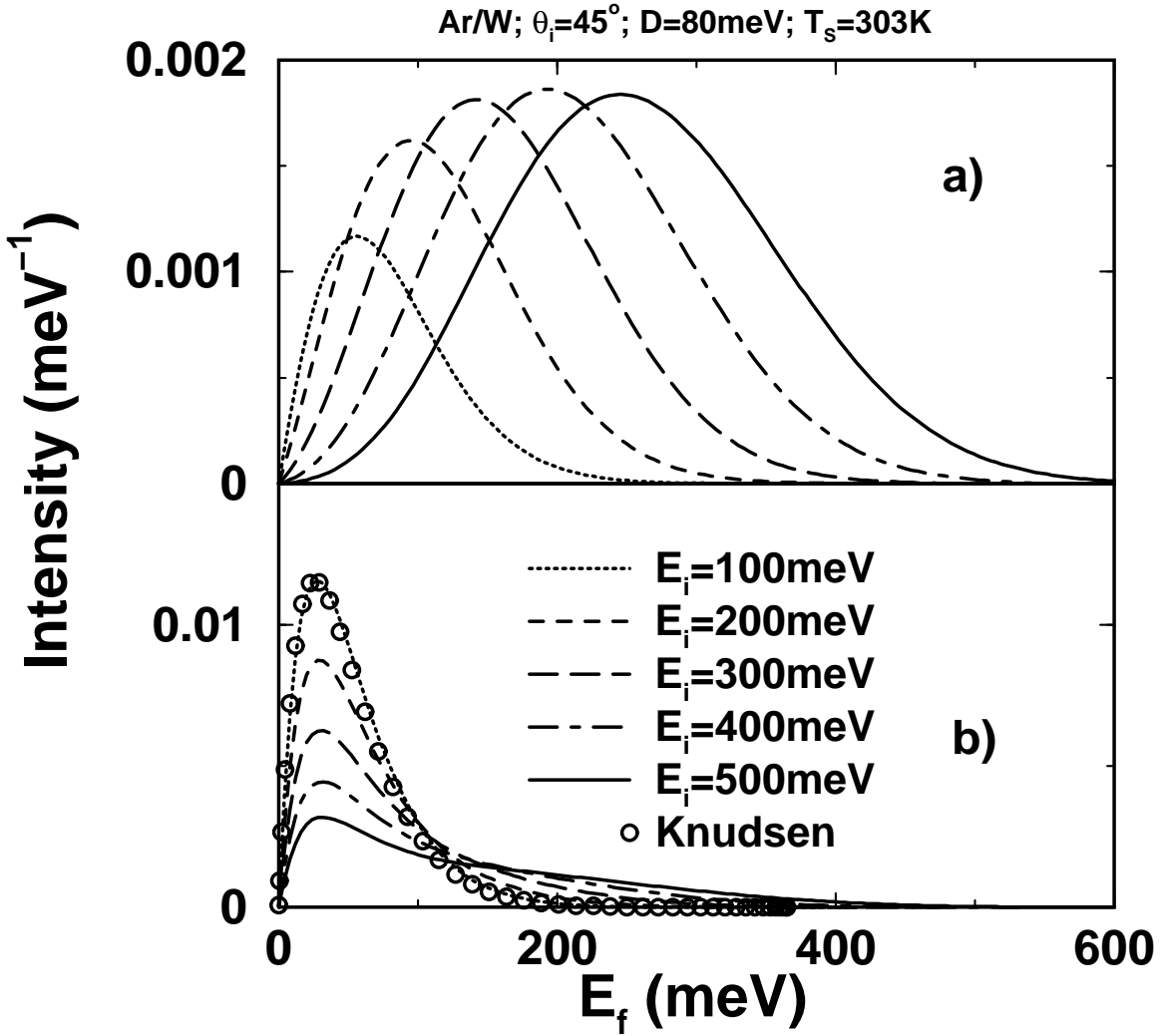}
\caption{The direct and trapping-desorption scattering energy distributions for incident energies large compared
to the well depth.  The system is Ar/W with $\theta_i=45^\circ$, $D=80$ meV, $T_S=303$ K and incident energies
as marked.  a) the upper panel shows the direct scattering contribution, and b) the lower panel shows the
trapping-desorption intensity.  A Knudsen distribution is also included in the lower panel.
}
\label{direct}
\end{figure}

\begin{figure}
\includegraphics[width=5.0in]{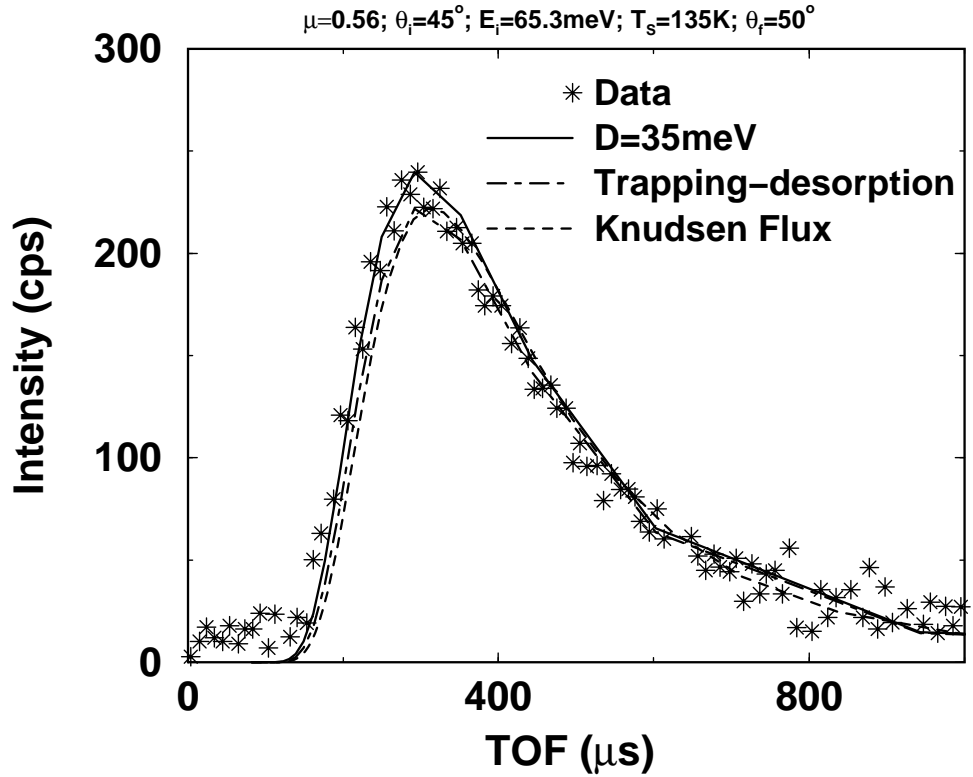}
\caption{
Intensity versus TOF for Ar scattering from a 1-decanethiol layer on Au(111) with $E_i=65.3$ meV, $\theta _i =45^\circ$ and $\theta _f =50^\circ$.  The calculation shown as a solid curve is the total differential reflection coefficient converted to TOF calculated with $\mu = 0.56$ and $D=35$ meV, the dash-dotted curve is the trapping-desorption fraction, and the dotted curve is a Knudsen equilibrium distribution.
}
\label{ei653}
\end{figure}

\begin{figure}
\includegraphics[width=5.0in]{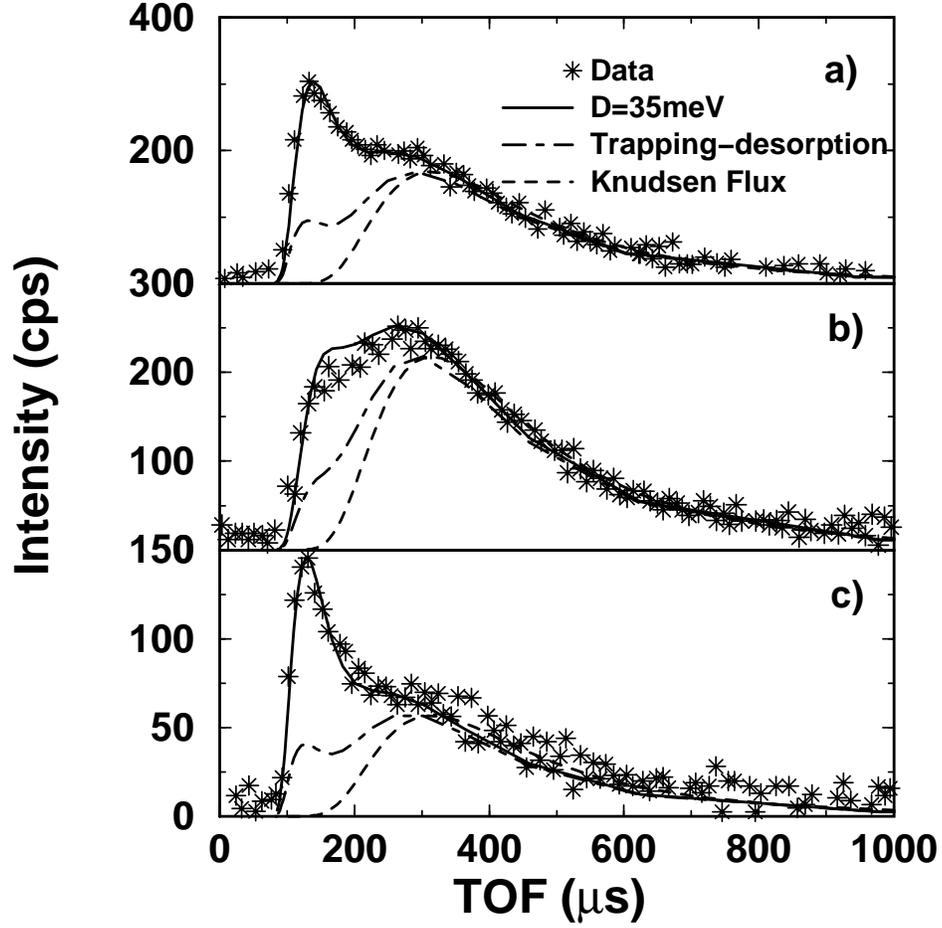}
\caption{
Intensity versus TOF for Ar scattering from a 1-decanethiol layer on Au(111) with $E_i=356$ meV: a)  $\theta _i =45^\circ$ and $\theta _f =50^\circ$, b) $\theta _i =30^\circ$ and $\theta _f =50^\circ$ and c) $\theta _i =30^\circ$ and $\theta _f =80^\circ$.  The solid curves are calculations with $\mu = 0.56$ and $D=35$ meV, the dash-dotted curves are the trapping-desorption fractions and the dotted curves are the Knudsen distribution. }
\label{ei365}
\end{figure}

\begin{figure}
\includegraphics[width=5.0in]{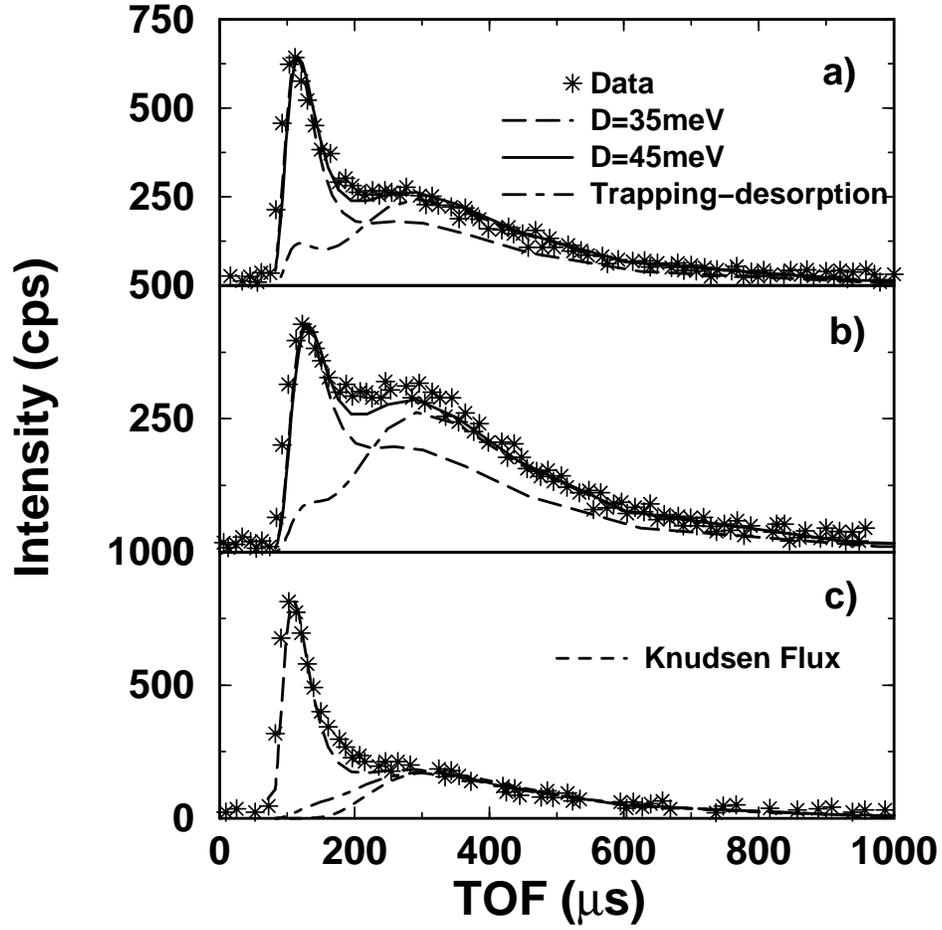}
\caption{
Intensity versus TOF for Ar scattering from a 1-decanethiol layer on Au(111) with $E_i=582$ meV: a)  $\theta _i =45^\circ$ and $\theta _f =50^\circ$,  b)  $\theta _i =45^\circ$ and $\theta _f =40^\circ$ and c)  $\theta _i =60^\circ$ and $\theta _f =40^\circ$.  The curves are labeled the same as in Fig.~\ref{ei365} except calculations are shown for both $D=35$ and 45 meV.
}
\label{ei582}
\end{figure}

\begin{figure}
\includegraphics[width=5.0in]{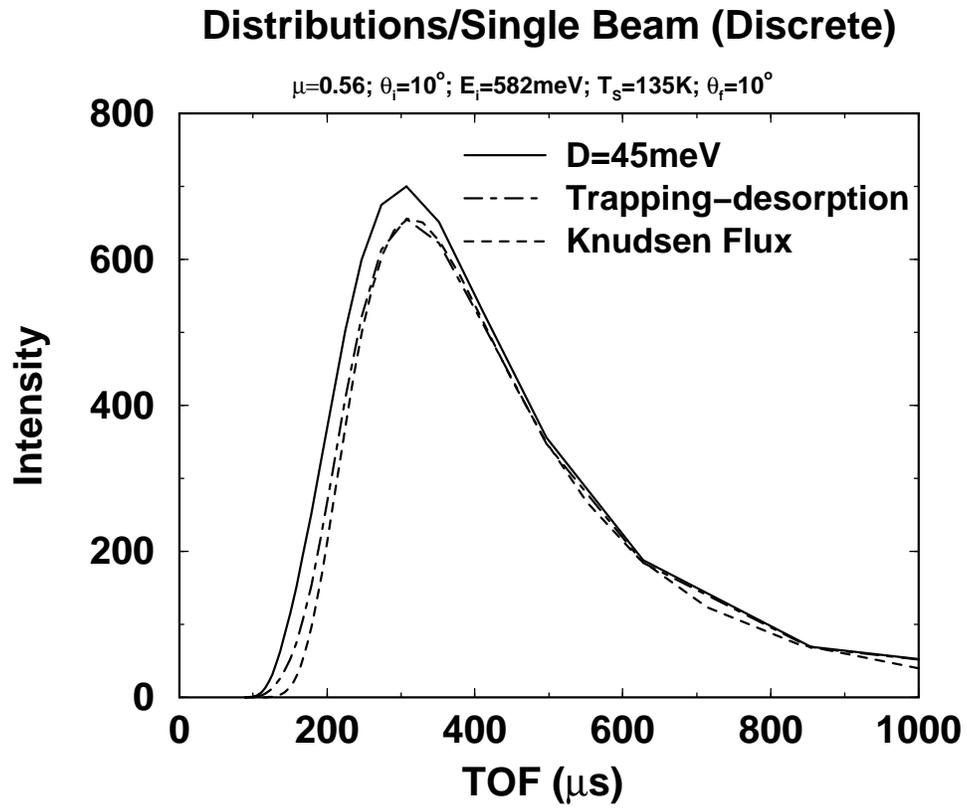}
\caption{
Calculations similar to those of Fig.~\ref{ei582} with $E_i=582$ meV and $D=35$ meV but with $\theta _i =10^\circ$ and $\theta _f =10^\circ$, showing that for normal incidence a distinct direct scattering peak is not expected.
}
\label{thf10}
\end{figure}

%XXXXXXXXXXXXXXXXXXXXXXXXXXXXXXXXXXXXXXXXXXXXXXXXXXXXXXXXXXXXXXXXXXXXXXXXXXXXXXXXXXXXXXXXXXXXXXXXX

\end{document}